\documentstyle[preprint,aps,epsf]{revtex}
\begin{document}
\tighten
\draft
\preprint{UG-30/99}
\title{
Nontriviality of Abelian gauged Nambu--Jona-Lasinio models 
in four dimensions}
\author{Manuel Reenders\footnote{Email address: m.reenders@phys.rug.nl}}
\address{
Institute for Theoretical Physics,\\
University of Groningen,
9747 AG Groningen, The Netherlands}
\date{April 2000}
\maketitle
\begin{abstract}
We study a particular class of Abelian gauged Nambu--Jona-Lasinio models
with global $U_L(N)\times U_R(N)$ symmetry, where $N$ is the number of 
fermion flavors.
We show, by treating the gauge interaction in
the ladder approximation and four-fermion interactions in the
leading order of the $1/N$ expansion, that 
the renormalization-group $\beta$ function of the $U(1)$ gauge coupling
has ultraviolet stable fixed points for sufficiently 
large $N$. This implies the existence 
of a nontrivial continuum limit.
\end{abstract}
\section{Introduction}
\begin{figure}[t!]
\epsfxsize=10cm
\epsffile[-40 400 330 540]{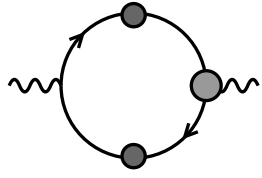}
\caption{The vacuum polarization tensor 
$\Pi_{\mu\nu}(q)=(-q^2g_{\mu\nu}+q_\mu q_\nu)\Pi(q^2)$, 
with the blobs representing full fermion propagators and a full 
photon-fermion vertex.}
\label{fig_vacpol}
\end{figure}

The absence of an interacting continuum limit or
triviality of Abelian gauge theories in four dimensions such as QED 
is due to the screening of charged particles
by their interactions with virtual fermion-antifermion pairs from
the vacuum. 
Such charge screening is described by the vacuum polarization $\Pi$ 
(Fig.~\ref{fig_vacpol}).
The QED vacuum is not a perfect insulator and can be considered as a medium 
of dipoles representing the fermion loops in the vacuum polarization.
Within perturbation theory, the electromagnetic charge or gauge coupling 
is screened completely in the continuum limit 
($\Lambda\rightarrow \infty$, where $\Lambda$ is the ultraviolet cutoff).

This can be seen by considering the 
renormalization group (RG) transformation of QED \cite{gemalo54}, which
relates the gauge coupling or fine-structure constant $\alpha_\mu$
in the infrared (IR) region to the bare gauge coupling
$\alpha_0=\alpha_\Lambda$ in the ultraviolet (UV) region
via 
\begin{eqnarray}
\alpha_\mu={\cal R}_{\alpha}\left[\mu/\Lambda,\alpha_0\right], \qquad 
{\cal R}_{\alpha}\left[\mu/\Lambda,\alpha_0\right]
=\frac{\alpha_0}{1+\Pi\left[\mu/\Lambda,\alpha_0\right]}.
\end{eqnarray}
In perturbation theory,
the one-loop leading contribution to $\Pi$ is 
\begin{eqnarray}
\Pi\left[\mu/\Lambda,\alpha_0\right]
=\frac{2N\alpha_0}{3\pi} \ln\frac{\Lambda}{\mu}, \label{vacpol1loop}
\end{eqnarray}
where $N$ is the number of fermion flavors.
This logarithmic screening effect is sufficient to 
cause the complete screening of charge in the continuum limit
\cite{lapo55,fr55}.

From the RG point of view triviality is merely due to  
the absence of an UV stable fixed point or nontrivial root of the 
$\beta$ function 
\begin{eqnarray}
\beta_\alpha(\alpha_0)\equiv 
\frac{\partial 
{\cal R}_{\alpha}\left[w,\alpha_0\right]}{\partial w}\Biggr|_{w=1}=
\Lambda \frac{{d}\alpha_0}{{d} \Lambda}
\label{betaadef}
\end{eqnarray}
for the gauge coupling.
Equation~(\ref{vacpol1loop}) gives rise to the 
$\beta$ function $\beta_\alpha(\alpha)=2N\alpha^2/3\pi$,
which only has an IR stable fixed point or trivial root: $\alpha=0$ 
($\beta_\alpha^\prime(0)\geq 0$).
A nontrivial or interacting theory only arises whenever $\beta_\alpha$ 
has a root which is an UV stable fixed point.
In case of asymptotic free theories such as QCD the origin $\alpha=0$
is an UV stable fixed point.

In addition to its motivation only on the level of perturbation theory
the above consideration is too naive.
According to the RG methods of Wilson 
\cite{wi71,wiko74}, one should consider the RG flow 
in the space of all coupling constants (respecting certain symmetries)
or at least for 
those coupling constants which can be classified as relevant close
to a particular critical manifold in coupling constant space.
For a possible nontrivial continuum limit of QED,
critical and hence nonperturbative dynamical effects are required.

Therefore, the discovery of dynamical chiral symmetry breaking (D$\chi$SB) in 
the strong coupling phase of QED ($\alpha_0\geq \alpha_c=\pi/3$) 
\cite{fuku76,fomi76,fogumisi76}
and the existence of an UV stable fixed point in the 
quenched-ladder approximation \cite{fogumi78,fogumisi83,mi85a} sheds new 
light on 
the nonperturbative nature of QED and triviality.       
Lattice simulations of so-called noncompact quenched QED have 
confirmed the existence of a continuous chiral phase transition  
\cite{kodako88a,kodako89b,kodako89c}.

An important step was performed by Bardeen, Leung, and Love
in Refs.~\cite{balelo86,leloba86}.
These authors realized that, in quenched-ladder QED, 
attractive four-fermion interactions described by a dimensionless 
four-fermion coupling $g_0\geq 0$ have a so-called scaling dimension 
4 instead of 6 at the critical gauge 
coupling $\alpha_0=\alpha_c$. Consequently, these operators mix with the gauge 
interaction which also has dimension 4 in four space-time dimensions;
this means that QED is not a closed theory at the chiral phase transition.
The scaling dimension of an operator is important in determining whether or 
not such an operator describes a long-range interaction.
The model of Bardeen, Leung, and Love is referred to as the gauged 
Nambu--Jona-Lasinio (GNJL) model, and the mechanism of dynamical chiral 
symmetry breaking proposed
by them should be regarded in the context of the RG methods of Wilson.

Wilson pointed out that a nontrivial renormalizable model can only be 
formulated if the RG transformation exhibits UV stable fixed points.
UV stable fixed points are ``singular'' points\footnote{
UV stable fixed points are specific roots of the $\beta$ functions.} 
of the RG transformation
at which the model becomes scale (conformal) invariant. 
Natural candidates for UV stable fixed points
are critical points governing a continuous phase transition.
Since, at a continuous phase transition, the correlation 
length $\xi$ is infinite and the model is scale invariant. 

The most crucial observation of Wilson is that in the RG transformation, 
{\em i.e.}, the ``coarse graining'' process, new types of local interactions
are generated and that the new interactions can be classified
as either irrelevant or relevant interactions.
Only relevant (including marginal) interactions are important in determining
what kind of infrared dynamics characterized by a scale $\mu\sim 1/\xi$ 
emerges from the microscopic or bare model characterized by the cutoff 
$\Lambda$.
The effect of irrelevant interactions can always be absorbed
by adapting the coupling constants of relevant and marginal interactions.

Especially, close to a continuous phase transition, 
the RG methods show that it is impossible, {\em a priori} (without solving the 
equations of motion), to determine which interactions are relevant or 
irrelevant;
particular interactions can acquire anomalous dimensions and 
interactions which are irrelevant 
in a certain region of coupling constant space
might become relevant in another region.

In this respect the GNJL model should be considered as the Wilsonian
effective (or microscopic) action of QED taking into account the four-fermion 
interactions describing neutral scalar and pseudoscalar 
fermion-antifermion composites. 
It was shown in Refs.~\cite{komiya89,apsotawij88} that there is a 
critical line (curve) in the coupling constant plane $(\alpha_0,g_0)$
of the GNJL model
separating a chiral symmetric phase from the chiral broken phase.
The critical line is given by
\begin{eqnarray}
g_c(\alpha_0)=(1+\omega)^2/4,\qquad \omega=\sqrt{1-4\lambda_0},
\qquad \lambda_0=3\alpha_0/4\pi.\label{critcurv2wdef}
\end{eqnarray} 
In the neighborhood of the critical line 
four-fermion interactions acquire sufficiently large anomalous dimensions 
to become relevant operators. 
The existence of nontrivial scaling behavior of the model near criticality
implies that these scalar and pseudoscalar composites 
are relevant degrees of freedom at both short and long 
distances.\footnote{The composites 
are light states, as well in the symmetric phase as in the broken phase.}

Lattice simulations of noncompact quenched QED with an induced four-fermion 
coupling were performed by the Illinois group in 
Refs.~\cite{kohakoda90,kokolowa93}.
The Illinois group obtained a critical point $(0.44 \alpha_c,0.76)$
in the $(\alpha_0,g_0)$ plane, which fits nicely on the critical line
Eq.~(\ref{critcurv2wdef}).
However it was not possible at that time to investigate by means of 
lattice simulations the phase transition along the critical line.

Later, the GNJL model was studied on the lattice in noncompact formulation 
using some mean-field approach for the fermions in 
Ref.~\cite{azcagagrlapi95b}.
They obtained a critical line qualitatively similar 
to the one following from the Schwinger--Dyson equation (SDE) approach 
Eq.~(\ref{critcurv2wdef}).

In the intermediate region ($0<\alpha_0< \alpha_c$),
the critical exponents describing the chiral phase transition 
satisfy nonmean-field hyperscaling relations 
which supports the view that, within the quenched-ladder mean-field 
approximation, the GNJL model has a nontrivial continuum 
limit~\cite{kohakoda90,balomi90}. Also in Ref.~\cite{azcagagrlapi95b} nonmean 
field critical exponents were obtained.

The physical implications and the consistency of the quenched-ladder results 
with many quenched lattice simulations, and with the nonperturbative RG 
techniques, support the view that the qualitative features
of the approach might be realistic and describe properties of the full theory.
Most likely this is due to the ladder approximation respecting the vector 
and axial Ward--Takahashi identities.

In Refs.\cite{aplama88,ho89} the validity of the ladder approximation
was tested positively 
by including the effects of, {\em e.g.}, crossed photon exchange graphs. 
In addition, the nonperturbative renormalization-group methods of 
Refs.~\cite{aomosuteto97,ao97} provide a way to
check the quenched-ladder approximation in the GNJL model
by including the effect of crossed photon exchange graphs, and four-fermion
interactions in the RG flow of coupling constants.
In Ref.~\cite{ao97} 
the critical line\footnote{They obtained the critical line: 
$g_c(\alpha_0)=\left(1+\omega\right)^2/3$, which differs by a factor $3/4$ 
from the quenched-ladder result.} 
and critical exponents in the full quenched GNJL model
were obtained in a particular so-called local potential approximation,
which incorporates besides crossed photon exchange graphs 
also four-fermion exchanges beyond the mean-field approach.
Considering the small quantitative differences, qualitatively
this study supports the reliability of the ladder approximation.

Attempts to include a logarithmic running of the coupling drastically 
changes the chiral phase transition and the critical line, 
see Refs.~\cite{gu90,go90,ra91,ko90}. 
Moreover, it was shown in Refs.~\cite{ko90,ko91,gu91,gosa91} that 
the critical exponents are of the mean-field type
(up to logarithmic violations) leading to a trivial theory. 

Lattice simulations of noncompact full (unquenched) QED on the lattice 
(with flavors, $N=2$ and $N=4$) are controversial \cite{goholirascst97}.
The Illinois group \cite{kokowa93,hakokoresiwa94} (see 
also \cite{kodako89a,kodako89b})
and the Zaragosa group \cite{azcagr93,azcagagrlapi95a}, 
find power-law scaling and nonmean-field critical exponents, signaling
a possible nontrivial continuum limit for the strong-coupling broken phase, 
whereas \cite{goholarascsowi90,gohorascso92,goholirascst97} 
obtain mean-field behavior (mean-field critical exponents 
with logarithmic corrections).
Thus G{\"o}ckeler {\em et al.} find a vanishing renormalized
gauge coupling and a vanishing effective Yukawa coupling (defined
by the Goldberger--Treiman relation), and they conclude 
that lattice QED is trivial, see for their most recent result
Ref.~\cite{goholirascst98}.

In this paper we argue that the four-fermion interactions 
might play a crucial role in the phenomenon of charge screening.
We show that, by considering the Abelian GNJL  
model, UV stable fixed points of the $\beta$ function of 
the gauge coupling can be realized,
provided the number of fermion flavors $N$ exceeds some 
critical value.
The existence of UV stable fixed points gives rise to a nontrivial theory. 

An important observation is that the nonmean-field values for the critical 
exponents suggest the existence of a nontrivial Yukawa interaction 
describing the interactions of the scalar and pseudoscalar
composites with fermions. This also 
points out the inconsistency of the mean-field approximation 
(Hartree--Fock approximation) for the four-fermion interactions.
Therefore, we go beyond the mean-field approximation by incorporating 
these composites (the $\sigma$ and $\pi$ exchanges)
in the $1/N$ expansion.

The setup of this paper is the following.
In Sec.~\ref{sec_AGNJL} we introduce the Abelian GNJL model with
$U_L(N)\times U_R(N)$ symmetry. 
Furthermore, we sketch how we will search for the existence of 
an UV stable fixed point
in the coupling constant plane $(\alpha_0,g_0)$.
In Sec.~\ref{sec_ren_hyp} we discuss, within the quenched-ladder mean-field
approximation, the importance of hyperscaling relations and how 
this is related to the existence of a nontrivial 
Yukawa interaction in the GNJL model.
We discuss how to proceed beyond the quenched mean-field approximation in
Sec.~\ref{sec_beyondqmf}.
The $1/N$ expansion is discussed in Sec.~\ref{ssec_cs1/N}.
In order to get some idea how the scalar and pseudoscalar composites 
contribute to vacuum polarization
we illustrate such contributions on the level of perturbation theory
for a gauge--Higgs--Yukawa model in Sec.~\ref{sec_gHY}.
Then, in Sec.~\ref{vacpol1n}, 
we argue how we can exploit the $1/N$ expansion and derive 
a computable expression for the $\beta$ function, $\beta_\alpha$, 
of the gauge coupling. 
Subsequently, $\beta_\alpha$ is computed explicitly in 
Sec.~\ref{sec_compbeta}.
The existence of UV stable fixed points is addressed in 
Sec.~\ref{critfixpoints}. 
Finally, we present our conclusions in Sec.~\ref{sec_concl}.
\section{The Abelian GNJL model}\label{sec_AGNJL}
We consider the GNJL model with $U(1)$ gauge symmetry
(the Abelian GNJL model) with $N$ number of fermion flavors described by 
the following Lagrangian (see also Refs.~\cite{mirbook,manuel}):
\begin{eqnarray}
{\cal L}_1&=&
\bar\psi_i (i\gamma^\mu D_{\mu}-m_0)\psi_i-\frac{1}{4}F_{\mu\nu}F^{\mu\nu}
+\frac{G_0}{2}\sum_{\alpha=0}^{N^2-1}\left[
\left(\bar \psi_i \tau^\alpha_{ij} \psi_j\right)^2
+\left(\bar\psi_i \tau^\alpha_{ij}(i{\gamma_5})\psi_j\right)^2\right],
\label{gnjlun}
\end{eqnarray}
where $D_{\mu}=\partial_{\mu}-ie_0A_{\mu}$ and where the flavor 
labels $i,j$ run from $1$ to $N$.
The Lagrangian ${\cal L}_1$ is parametrized by the following 
three dimensionless bare coupling constants:
\begin{eqnarray}
\mu_0=m_0/\Lambda\equiv m_\Lambda,\qquad
\alpha_0=e_0^2/4\pi\equiv \alpha_\Lambda,\qquad 
g_0=G_0\Lambda^2/4\pi^2\equiv g_\Lambda,
\end{eqnarray}
where $\Lambda$ is the ultraviolet cutoff.
We assume that the above set of dimensionless coupling constants
comprises the entire set of relevant (including marginal) 
dimensionless coupling constants respecting particular chiral, vector, 
and gauge symmetries.

If the bare mass $m_0$ is zero, ${\cal L}_1$ has a global 
$U_L(N)\times U_R(N)$ symmetry. The generators, $\tau$, of the $U(N)$ Lie 
algebra have the following properties:
\begin{eqnarray}
\tau^{\alpha\dagger}=\tau^\alpha,\qquad
{\rm Tr}\, \tau^\alpha \tau^\beta=\delta^{\alpha\beta},\qquad
\sum_{\alpha=0}^{N^2-1}\tau^\alpha_{ij} \tau^\alpha_{kl}
=\delta_{il}\delta_{kj}.\label{Fierz}
\end{eqnarray} 
The last identity is called the Fierz identity.

The Abelian GNJL model described by Eq.~(\ref{gnjlun}) can be conveniently 
analyzed in terms of auxiliary or composite fields 
$\sigma^\alpha=-G_0\bar\psi\tau^\alpha\psi$ and 
$\pi^\alpha=-G_0\bar\psi\tau^\alpha(i{\gamma_5})\psi$ describing scalar and 
pseudoscalar degrees of freedom.
In this way ${\cal L}_1$ can be rewritten as
\begin{eqnarray}
{\cal L}_2&=&\bar\psi_i i\gamma^\mu D_\mu\psi_i
-\frac{1}{4}F_{\mu\nu}F^{\mu\nu}-\sum_{\alpha=0}^{N^2-1}
\bar\psi_i\tau^\alpha_{ij}(\sigma^\alpha+i{\gamma_5}\pi^\alpha)\psi_j
-\frac{1}{2G_0}\sum_{\alpha=0}^{N^2-1}\left[ (\sigma^\alpha)^2
+(\pi^\alpha)^2\right],\label{unaux}
\end{eqnarray}
where $m_0$ has been set to zero.
In this formulation the four-fermion interactions are described 
by the interactions of the auxiliary fields with the fermion fields.
Then the connected two-point Green functions of the $\pi^\alpha$ 
fields describe
$N^2$ Nambu--Goldstone bosons ($\pi$ bosons) and the connected two-point 
Green functions of the $\sigma^\alpha$ fields describe $N^2$ ``Higgs'' bosons 
($\sigma$ bosons).

Although the Lagrangian (\ref{gnjlun}) has very interesting properties,
surprisingly this particular class of GNJL models with
$U_L(N)\times U_R(N)$ has not received much attention. 
One such property is that Eq.~(\ref{gnjlun}) 
comprises the largest set of relevant chiral invariant four-fermion 
operators for a model of $N$ fermions. 
Another independent set of vectorlike chiral invariant  
four-fermion interactions such as $(\bar\psi \gamma^\mu \psi)^2$
do not acquire large anomalous dimensions and remain irrelevant
near the critical point (line).
This has been shown in Refs.~\cite{aomosuteto97,ao97}, where the RG flow
of scalarlike chiral invariant and vectorlike chiral invariant
four-fermion interactions has been considered. 
Moreover, the specific form of the chiral symmetry, where 
the number of scalars equals the number of pseudoscalars,
turns out to have useful implications in the context of $1/N$ expansions 
as we will discuss later in Sec.~\ref{sec_beyondqmf}.

In case of ${\cal L}_2$ there are four renormalization constants:
\begin{eqnarray}
&&Z_2^{1/2}(\Lambda'/\Lambda)\psi_{(\Lambda')}(x)=\psi(x), \quad 
Z_3^{1/2}(\Lambda'/\Lambda)A^\mu_{(\Lambda')}(x)=A^\mu(x),\label{Z1Z2Z3def}\\
&&Z_\sigma(\Lambda'/\Lambda) \sigma_{(\Lambda')}(x)=\sigma(x),\quad
Z_\pi(\Lambda'/\Lambda) \pi_{(\Lambda')}(x)=\pi(x), \label{Zsigmapidef}
\end{eqnarray} 
where $\Lambda'/\Lambda\leq 1$, and the fields $\psi$, $A^\mu$, $\pi$, 
$\sigma$ are the bare fields defined at the UV cutoff $\Lambda$. 

Near the critical point a new scale is generated: the correlation 
length $\xi$.
In case of a second-order type of phase transition,
the correlation length exists in both phases of the system. 
In the broken phase the inverse correlation length is real and can be 
considered as a physical mass of particles, for instance the mass of 
the scalar bound state (the $\sigma$ boson) $m_\sigma$, 
or the mass of fermion $m_{\rm dyn}$. 
In the symmetric phase the fermion is massless, and the
scalar and pseudoscalar composites are unstable states characterized by a  
complex mass pole in their respective propagators describing the
mass and the width of the Breit--Wigner type resonance,
see Refs.~\cite{aptewij91,hakoko93,gure98}.
The absolute value of the complex mass pole $|m_\sigma|$ can be considered
as the inverse correlation length, {\em i.e.}, $|m_\sigma|\sim 1/\xi$.

The RG transformation dictates the flow of 
the dimensionless bare couplings  as function of the UV cutoff $\Lambda$.
Typically the bare relevant and marginal couplings 
({\em e.g.}, $\mu_0$, $\alpha_0$, $g_0$) have to be fine-tuned sufficiently 
close to the critical point in order for scaling behavior to set in,
so that the physics in the infrared can be related to experimental data. 
Scaling behavior is obtained when there is a large scale hierarchy 
between the infrared length scale $\xi$ and the ultraviolet length 
scale $a=1/\Lambda$, {\em i.e.}, $\xi\gg a$.

The fine-tuning depends on the eigenvalues of the RG 
transformation of the couplings close to the critical point and hence on the
critical exponents.
These critical exponents can be derived from 
the $\beta$ functions for the coupling constants 
$\mu_0$, $\alpha_0$, and $g_0$; 
\begin{eqnarray} 
\Lambda\frac{{d} \mu_0}{{d} \Lambda}=\beta_\mu(\mu_0,\alpha_0,g_0),\quad
\Lambda\frac{{d} \alpha_0}{{d} \Lambda}=\beta_\alpha(\mu_0,\alpha_0,g_0),\quad
\Lambda\frac{{d} g_0}{{d} \Lambda}=\beta_g(\mu_0,\alpha_0,g_0).
\label{betamug0a0}
\end{eqnarray}
The crucial step is determine the fixed points 
$(\mu_\star,\alpha_\star,g_\star)$ of the RG equations (\ref{betamug0a0}), 
{\em i.e.},
\begin{eqnarray}
\beta_\mu(\mu_\star,\alpha_\star,g_\star)=0,\quad
\beta_\alpha(\mu_\star,\alpha_\star,g_\star)=0,\quad
\beta_g(\mu_\star,\alpha_\star,g_\star)=0,
\end{eqnarray}
since the nature of the fixed point determines whether a nontrivial 
continuum limit can be realized or not.
For a nontrivial continuum limit $(\mu_\star,\alpha_\star,g_\star)$
should be a UV stable fixed point.

The RG equations follow from the regularized SDE's of the generating 
functional described by the Lagrangian (\ref{unaux}).\footnote{A derivation 
of the set of (full) SDE's for the two and three-point functions is given 
in Chap.~2 of \cite{manuel}.}   
The UV stable fixed point for $\beta_\mu$ is $\mu_\star=0$,
hence we can write
\begin{eqnarray}
\Lambda\frac{{d} \mu_0}{{d} \Lambda}\approx-\left(1+\gamma_m\right) \mu_0,
\label{gammamdef}
\end{eqnarray}
where $\gamma_m$ is the anomalous dimension 
of the mass operator $\bar\psi\psi$ evaluated at the fixed point 
$(\mu_\star,\alpha_\star,g_\star)$.
In the case of the quenched GNJL models $1\leq \gamma_m \leq 2$.
Thus the dimensionless bare mass $\mu_0$ is a relevant coupling requiring
fine-tuning.

After setting $\mu_0=\mu_\star=0$,
the problem reduces to the determination of the UV stable fixed points 
in the coupling constant plane $(\alpha_0,g_0)$: {\em i.e.},
\begin{eqnarray}
\beta_g(\alpha_\star,g_\star)&=&0,\qquad 
\beta^\prime_g(\alpha_\star,g_\star)=
\frac{\partial \beta_g(\alpha,g)}{\partial g}\biggr|_{
(\alpha,g)=(\alpha_\star,g_\star)}<0,\label{subbetag0}\\
\beta_\alpha(\alpha_\star,g_\star)&=&0,\qquad
\beta^\prime_\alpha(\alpha_\star,g_\star)=
\frac{\partial \beta_\alpha(\alpha,g)}{\partial \alpha}\biggr|_{
(\alpha,g)=(\alpha_\star,g_\star)}<0.
\label{UVfixpointdef}
\end{eqnarray}

The quenched-ladder approximation
simplifies the solutions of Eqs.~(\ref{subbetag0}) and 
(\ref{UVfixpointdef}) considerably 
since the quenched hypothesis explicitly sets $\beta_\alpha=0$
for all $\alpha_0$ by omiting fermion loops.
It was shown in Ref.~\cite{kotaya93} that, in the symmetric phase 
($g_0\leq g_c$),
\begin{eqnarray}
\beta_g(\alpha_0,g_0)=-2\omega\frac{g_0}{g_c}(g_0-g_c), \label{RGg0sym} 
\end{eqnarray}
with $\omega$ and $g_c$ given in Eq.~(\ref{critcurv2wdef}).
In the next section, Eq.~(\ref{RGg0sym}) will be derived.
Clearly, in this way, the UV stable fixed point of $\beta_g$
is the critical line;
\begin{eqnarray}
g_\star=g_c(\alpha_0),\qquad \beta_g(\alpha_0,g_c(\alpha_0))=0.
\end{eqnarray}
Now Eq.~(\ref{UVfixpointdef}) should be reconsidered.
We will analyze Eq.~(\ref{UVfixpointdef}) beyond the quenched approximation, 
and try to solve 
\begin{eqnarray}
\beta_\alpha(\alpha_0,g_c(\alpha_0))=0. \label{betaa0gc}
\end{eqnarray}
In Sec.~\ref{vacpol1n} an explicit expression for $\beta_\alpha$ 
will be derived by assuming that $g_0$ is at its critical value $g_c$ and
that it has an UV stable fixed point so that in the neighborhood of this point
$\beta_\alpha\approx 0$. 
\section{Hyperscaling in the quenched-ladder mean-field 
approximation}\label{sec_ren_hyp}
In analogy with statistical mechanics, the continuous chiral phase transition 
can be classified in terms of critical exponents which describe the scaling of
various macroscopic quantities ({\em e.g.}, the chiral condensate, 
correlation length, effective potential, 
chiral susceptibility) close to or at the critical point.
It is considered a strong indication of the existence of a nontrivial 
continuum limit ($\Lambda\rightarrow \infty$), 
if so-called hyperscaling relations between these various critical exponents 
are satisfied, see Refs.~\cite{kohakoda90,balomi90,amit,goldenfeld,zinjus}.

Because the $\sigma$ boson propagator $\Delta_{S}$ 
is the connected correlation function of the field $\sigma$ 
describing correlations parallel to direction of symmetry breaking 
({\em i.e.}, parallel to the direction 
of long-range ordering),
the absolute value of the mass, $m_\sigma$, of the $\sigma$ boson,
which is given by $\Delta_S$, is the natural
candidate for the inverse correlation length.

In the quenched-ladder approximation treating four-fermion interaction
in a mean-field approximation, 
the critical exponents are
\begin{eqnarray}
\delta=\frac{2+\omega}{2-\omega},\qquad 
\beta=\frac{2-\omega}{2\omega},\qquad
\nu=\frac{1}{2\omega},\qquad \gamma=1,\label{delbetgamnu}
\end{eqnarray}
and satisfy the hyperscaling relations
\begin{eqnarray}
\gamma=\beta(\delta-1),\qquad 4\nu =2\beta+\gamma.\label{hyperscal}
\end{eqnarray}
Other hyperscaling relations involving the critical exponent $\alpha$ 
describing the scaling of the effective potential are satisfied too
\cite{kohakoda90,balomi90}.
Furthermore, it was argued in \cite{kohakoda90} that the validity of the 
quenched-ladder mean-field approximation relies on the verification
that the critical exponent $\gamma=1$.
The interpretation of $\gamma=1$ is the factorization
$\eta_{(\bar\psi\psi)^2}=2\eta_{(\bar\psi\psi)}$.
A renormalization of the chiral condensate simultaneously renormalizes
the propagators $\Delta_{S}$ and $\Delta_{P}$.
Indeed the lattice computations of the critical exponent $\gamma$ 
reported in \cite{kohakoda90,kokolowa93,azcagagrlapi95b} showed strong 
evidence for $\gamma=1$.

\begin{figure}[t!]
\epsfxsize=10cm
\epsffile[-40 400 330 540]{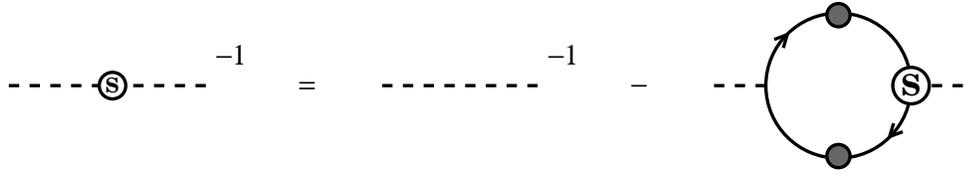}
\caption{SDE for the scalar propagator or $\sigma$ boson $\Delta_{S}(q)$.}
\label{fig_scalar}
\end{figure}

\begin{figure}[t!]
\epsfxsize=10cm
\epsffile[-40 400 330 540]{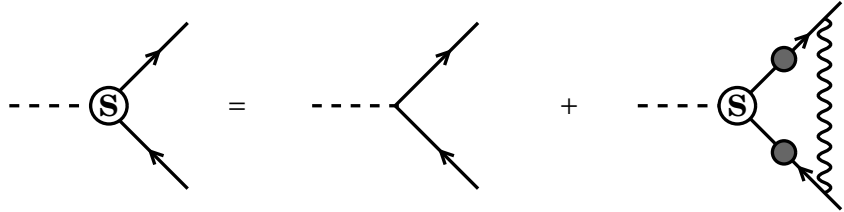}
\caption{SDE for the scalar vertex or Yukawa vertex ${\Gamma_{S}}(p+q,p)$ in 
the quenched-ladder approximation.}
\label{fig_scalarvertexladder}
\end{figure}

The anomalous dimension $\eta$ describes the scaling of the connected 
two-point Green function $\Delta_{S}$ at the critical point.
In Ref.~\cite{gure98} the scalar propagator $\Delta_{S}$ 
(see Fig.~\ref{fig_scalar}) and the scalar Yukawa 
vertex ${\Gamma_{S}}$ (see Fig.~\ref{fig_scalarvertexladder}) 
have been computed in the symmetric phase in the 
quenched-ladder approximation by means of a so-called
two-channel approximation.

In the symmetric phase the Yukawa vertex has the following form:
\begin{eqnarray}
{\Gamma_{S}}(p+q,p)= {\bf 1}\left[ F_1(p+q,p)
+\left({\hat q}{\hat p}-{\hat p}{\hat q}\right)F_2(p+q,p)\right]. 
\label{vertfiesdef}
\end{eqnarray}
These vertex functions $F_1$ and $F_2$ can be expanded in terms of 
Chebyshev polynomials of the second kind, {\em e.g.},
\begin{eqnarray}
F_1(p+q,p)&=&\sum_{n=0}^\infty f_n(p^2,q^2) U_n(\cos\alpha),\qquad
\cos \alpha=\frac{p\cdot q}{p q}.\label{chebF1}
\end{eqnarray}
The two-channel approximation of Ref.~\cite{gure98} now holds in
that the Yukawa vertex is approximated by the angular average 
of the vertex function $F_1$ in the following way:
\begin{eqnarray}
{\Gamma_{S}}(p+q,p)\approx {\bf 1} \int\frac{{d}\Omega_p}{2\pi^2} F_1(p+q,p)
={\bf 1} f_0(p^2,q^2), \label{canonic}
\end{eqnarray}
where
\begin{eqnarray}
f_0(p^2,q^2)\equiv F_{\rm IR}(p^2,q^2)\theta(q^2-p^2)
+ F_{\rm UV}(p^2,q^2)\theta(p^2-q^2).\label{channelapprox}
\end{eqnarray}
The functions $F_{\rm IR}$ 
and $F_{\rm UV}$ are, respectively, referred to as the IR channel
(infrared), and the UV channel (ultraviolet).
The specific choice of Chebyshev expansion 
and consequently, the choice of zeroth-order coefficient 
$f_0$, is convenient, since the infrared limit ($q^2\gg p^2$), 
and ultraviolet limit ($q^2\ll p^2$) of ${\Gamma_{S}}$ are both 
described by $f_0$.
The scaling form for $\Delta_{S}(q)$ is well described by these two 
functions $F_{\rm IR}$ and $F_{\rm UV}$.

The zeroth-order Chebyshev expansion or two-channel approximation 
gives second-order differential equations for the lowest order harmonic
with appropriate IR and UV boundary conditions.
These differential equations are exactly solvable and
the solutions are expressed in terms of a Bessel function of the first kind 
for $F_{\rm IR}$
and in terms of modified Bessel functions for $F_{\rm UV}$. 
The solutions are given in \cite{gure98}.

With the ``asymptotic solutions'' for ${\Gamma_{S}}$ given in terms of
$F_{\rm IR}$ and $F_{\rm UV}$, 
an analytic expression for the $\sigma$ boson 
can be obtained from
\begin{eqnarray}
\Pi_{S}(q^2)=\frac{\Lambda^2}{4\pi^2}\frac{1}{\lambda_0}
\left[ F_{\rm UV}(\Lambda^2,q^2)-1\right], 
\qquad \Delta_{S}^{-1}(q)=-\frac{1}{G_0}+\Pi_{S}(q^2),
\label{analyticscal}
\end{eqnarray}
which has been argued in \cite{gure98} to be correct up to 
leading and next-to-leading order in $q^2/\Lambda^2$ and is 
valid along the entire critical curve $g_c$.

Also one can derive that the solutions for $\Delta_{S}$ and
are consistent with hyperscaling for $0<\alpha_0<\alpha_c$.
This is intimately related to the fact that
the renormalization of the auxiliary fields 
$\sigma$ and $\pi$, Eq.~(\ref{Zsigmapidef}), simultaneously renormalizes 
the Yukawa vertex and the scalar propagator.

Near $g_0=g_c$ (with $|m_\sigma|^2, q^2 \ll \Lambda^2$)
the scalar propagator $\Delta_{S}$ has the scaling form
\cite{manuel,amit,goldenfeld,zinjus} (in Euclidean formulation):
\begin{eqnarray}
\Delta_{S}(q)=\frac{1}{\Lambda^2}
\left(\frac{\Lambda^2}{q^2}\right)^{1-\eta/2} 
{\cal F}_\Delta (|m_\sigma|^2/q^2),\qquad 
{\cal F}_\Delta (x)\approx -\frac{4\pi^2}{B(\omega)}\frac{1}{1+x^\omega},
\label{scalhypscalprop}
\end{eqnarray}
where $\eta$ is the anomalous dimension
\begin{eqnarray}
\eta=2(1-\omega),\label{etaexpr}
\end{eqnarray}
and where
\begin{eqnarray}
B(\omega)&\equiv&\frac{16 \omega}{(1-\omega^2)^2}
\frac{\gamma(-\omega)}{\gamma(\omega)} 
\frac{\Gamma(2-\omega)}{\Gamma(2+\omega)}
\left(\frac{\lambda_0}{2}\right)^\omega,\label{Bw}\\
\gamma(\omega)&\equiv&\sqrt{2\lambda_0}\left[
J_1(\sqrt{2\lambda_0})I_\omega^\prime(\sqrt{2\lambda_0})
+J_1^\prime(\sqrt{2\lambda_0})I_\omega(\sqrt{2\lambda_0})\right].
\label{gammaeq}
\end{eqnarray}
With Eq.~(\ref{delbetgamnu}), the anomalous dimension $\eta$, 
Eq.~(\ref{etaexpr}), satisfies the hyperscaling relation
\begin{eqnarray}
\gamma=\nu(2-\eta).
\end{eqnarray}

It has been shown in \cite{aptewij91,gure98} 
that $\Delta_{S}$ has a complex pole on a second 
Riemann sheet in (Minkowsky) momentum space.
The mass $m_\sigma$ is the complex pole of $\Delta_{S}$ and the absolute 
value $|m_\sigma|$ scales according to $|m_\sigma| \sim (\Delta g_0)^\nu$. 
More precisely, in \cite{aptewij91,gure98} it is derived that
\begin{eqnarray}
|m_\sigma|\propto \Lambda\left[
\frac{-\Delta g_0}{g_c g_0 B(\omega)}\right]^{1/2\omega},\qquad
\Delta g_0=g_0-g_c.
\label{resonance}
\end{eqnarray}
The absolute value of $m_\sigma$ is taken to be the physical, 
macroscopic or infrared mass scale, 
which by definition should be independent of $\Lambda$.
Using Eq.~(\ref{resonance}), we can derive the $\beta$ function of $g_0$ 
by assuming that
\begin{eqnarray}
0=\Lambda \frac{{d} |m_\sigma|}{{d}\Lambda}
\,\Longrightarrow \, \beta_g(\alpha_0,g_0)=-2\omega\frac{g_0}{g_c}(g_0-g_c),
\end{eqnarray}
which is equivalent to the $\beta$ function given in Ref.~\cite{kotaya93} 
and Sec.~\ref{sec_AGNJL}.
Hence the critical curve $g_0=g_c$ is an UV stable fixed point 
$\beta_g(\alpha_0,g_c)=0$
($\beta^\prime_g(\alpha_0,g_c)< 0$) of the RG flow.

In accordance with Ref.~\cite{gure98},
the scaling form for the Yukawa vertex can be written as
\begin{eqnarray}
{\Gamma_{S}}(p+q,p)\approx {\bf 1}\left(\frac{\Lambda^2}{q^2}\right)^{\eta/4}
\left[{\cal F}_{\rm IR}\left(p^2/q^2\right)\theta(q^2-p^2)
+{\cal F}_{\rm UV}\left(q^2/p^2\right)\theta(p^2-q^2)\right],
\label{scalformyukawa}
\end{eqnarray}
where, for $p^2,\,q^2 \ll \Lambda^2$,
\begin{eqnarray}
F_{\rm IR}(p^2,q^2)\approx \left(\frac{\Lambda^2}{q^2}\right)^{\eta/4}
{\cal F}_{\rm IR}\left(p^2/q^2\right),\qquad 
F_{\rm UV}(p^2,q^2)\approx 
\left(\frac{\Lambda^2}{q^2}\right)^{\eta/4}
{\cal F}_{\rm UV}\left(q^2/p^2\right),\label{scalfuv1}
\end{eqnarray}
and
\begin{eqnarray}
{\cal F}_{\rm IR}\left(p^2/q^2\right)&=&\frac{2\sin\omega\pi}{\pi}
\frac{2}{\gamma(\omega)}\frac{\Gamma(1-\omega)}{(1+\omega)}
\left(\frac{\lambda_0}{2}\right)^{\omega/2}\left(\frac{q^2}{p^2}\right)^{1/2}
J_1\left(\sqrt{\frac{2\lambda_0 p^2}{q^2} }\right)
,\label{scalfir2}\\
{\cal F}_{\rm UV}\left(q^2/p^2\right)&=&
\frac{2}{\gamma(\omega)}\frac{\Gamma(1-\omega)}{(1+\omega)}
\left(\frac{\lambda_0}{2}\right)^{\omega/2}
\left(\frac{q^2}{p^2}\right)^{1/2}\nonumber\\&\times&
\left[\gamma(\omega)I_{-\omega}\left(\sqrt{\frac{2\lambda_0 q^2}{p^2}}\right)-
\gamma(-\omega)I_{\omega}\left(\sqrt{\frac{2\lambda_0 q^2}{p^2}}\right)
\right].\label{scalfuv2}
\end{eqnarray}
From the above scaling form for $\Delta_{S}$ and ${\Gamma_{S}}$ it is clear 
that four-fermion scattering amplitudes such as
\begin{eqnarray}
{\Gamma_{S}}(p_1+q,p_1)\Delta_{S}(q){\Gamma_{S}}(p_2,p_2+q)
\propto \frac{1}{q^2},\qquad p_1^2,\, p_2^2 \ll q^2 \ll \Lambda^2,
\label{longnatYF}
\end{eqnarray}
are independent of $\Lambda$ and express 
the long-range nature\footnote{The correlation length is large,
$1/|m_\sigma|\gg 1/\Lambda$.} of Yukawa forces. 
For dimensions $2<d<4$, this was pointed out in Ref.~\cite{hakoko93}.
The long-range Yukawa forces and their nontrivial contributions 
to scattering amplitudes in the infrared are a direct consequence of 
hyperscaling and thus powerlike renormalizability.

The consensus is that in four dimensions due to the logarithmic corrections, 
the hyperscaling relations are violated for the pure NJL model and
$\lambda\phi^4$ theory, 
and we have the following inequalities:
\begin{eqnarray}
4\nu> 2\beta +\gamma,\qquad \gamma > \beta(\delta-1),\label{scallawineq}
\end{eqnarray}
see Ref.~\cite{koko94} for an extensive discussion regarding this issue.
This violation of hyperscaling is believed to be a sign of triviality
meaning that the effective Yukawa coupling (which couples
Goldstone bosons to the fermions) vanishes in the continuum limit.
The continuum limit is noninteracting, hence trivial.
This can be seen in the following way. 
Assuming that in the low-energy region the correlation length $\xi$  
is the only relevant length scale, we define an effective Yukawa coupling 
$g_Y$ by the zero-momentum limit of the four-fermion scattering amplitude 
(two fermions exchanging a scalar bound state) in the D$\chi$SB phase
\begin{eqnarray}
\frac{g_Y^2}{m_\sigma^2} \sim \xi^2 g_Y^2\sim  
{\Gamma_{S}}(0,0)\Delta_{S}(0){\Gamma_{S}}(0,0),
\end{eqnarray}
where ${\Gamma_{S}}(0,0)$ and $\Delta_{S}(0)$ are given by the chiral 
susceptibility relations 
\begin{eqnarray}
\Delta_{S}(0)=-G_0\frac{\partial \langle \sigma \rangle}{\partial m_0},
\qquad\Gamma_{S}(p,p)=({\bf 1}) 
\frac{\partial \Sigma(p^2)}{\partial \langle \sigma \rangle},\label{scalanal}
\end{eqnarray}
where $\Sigma_0=\Sigma(0)$ is the fermion mass in the broken phase.
By making use of the scaling laws \cite{kohakoda90,balomi90}
and that $\xi\sim 1/\Sigma_0\sim 1/m_\sigma$, 
$\langle \sigma \rangle \sim \langle \bar\psi\psi\rangle$, 
we can derive that 
\begin{eqnarray}
g_Y^2 \sim \xi^{(2\beta+\gamma-4\nu)/\nu}. \label{scallawyukawa}
\end{eqnarray}
This expression is related to the definition of 
$g_R\sim g_Y^2$ given in Ref.~\cite{koko94}, and it is clear that 
the scaling inequalities (\ref{scallawineq}) imply that $g_Y^2\rightarrow 0$ 
when $\xi\rightarrow \infty$.
Thus, only if the hyperscaling relations (\ref{hyperscal}) are satisfied, a 
nonzero $g_Y$ might be realized in the continuum limit 
($\xi\rightarrow \infty$),
thereby giving rise to a nontrivial interacting theory.
\section{Beyond the quenched-ladder mean-field 
approximation}\label{sec_beyondqmf}
The quenched approximation is analogous to the assumption that the full 
photon propagator 
\begin{eqnarray}
D_{\mu\nu}(q)=\left(-g_{\mu\nu}+\frac{q_\mu q_\nu}{q^2}\right)
\Delta(q)-a\frac{q_\mu q_\nu}{q^4},
\qquad \Delta(q)\equiv \frac{1}{q^2}\frac{1}{1+\Pi(q^2)}
\label{bargb}
\end{eqnarray}
can be approximated by the bare or canonical propagator $\Delta(q)=1/q^2$
(for large momenta), 
\begin{eqnarray}
D_{\mu\nu}(q)=\left(-g_{\mu\nu}+\frac{q_\mu q_\nu}{q^2}\right)\frac{1}{q^2}, 
\label{canphot}
\end{eqnarray}
in the Landau gauge ($a=0$). 
The quenched approximation is only consistent when the vacuum polarization 
is finite in the continuum limit, {\em i.e.}, 
the logarithmic running of the coupling is absent.
This is the case at an UV stable fixed point of the $\beta$ function,
Eq.~(\ref{UVfixpointdef}), of $\alpha_0$.
The assumption that such a critical fixed point exists, and that it lies 
somewhere on the critical
curve (\ref{critcurv2wdef}) is the starting point for many studies 
of dynamical chiral symmetry 
breaking in context of the GNJL model. 
In fact, the quenched hypothesis (\ref{canphot}) is only consistent 
when the bare gauge coupling $\alpha_0$ is near the fixed point of 
the theory, $\beta_\alpha\approx 0$.
We discuss this issue in more detail in Sec.~\ref{vacpol1n}.

In many approximations of the GNJL model, the four-fermion interactions are 
treated in a mean-field approach known as the Hartree--Fock approximation.
In mean-field approximations the 
composite operators such as $\bar\psi\psi$ are replaced 
by their vacuum expectation values 
($\bar\psi\psi\rightarrow\langle\bar\psi\psi\rangle$) and fluctuations about 
that value are ignored.
Thus quantum corrections corresponding to four-fermion interactions are 
neglected beyond tree level.

As long as four-fermion interactions are irrelevant the mean-field
approach for these operators is justified.
However, in Refs.\cite{kohakoda90,balomi90} it is concluded that, 
within the quenched-ladder mean-field approximation to the GNJL model,
the hyperscaling equations for the critical exponents are satisfied, 
implying that the four-fermion operators become relevant due to 
the appearance of large anomalous dimensions.
In other words, the mean-field approach
yields non-mean-field exponents, thereby being inconsistent
({\em e.g.}, see Refs.\cite{azcagagrlapi95b,goldenfeld}).
As was discussed in Sec.~\ref{sec_ren_hyp},
the hyperscaling relations imply the existence of a nontrivial
Yukawa interaction describing the interaction
between fermions and $\sigma$ and $\pi$ composites
in the GNJL model at both short and long distances.

In order to go beyond the mean-field approach, we propose the following.
First, we point out the usefulness of skeleton expansions.
Second, we make use of the specific form of the chiral symmetry
and adopt the $1/N$ expansion (with $N$ the number of fermion flavors).
\subsection{The skeleton expansion}\label{ssec_skel}
The non-mean-field values of the critical exponents imply 
that one cannot neglect (as is done in mean-field approximations)
the full connected Green functions corresponding to the composites  
(or at least the leading or asymptotic parts of these functions) in the SDE's.

On the level of the Bethe--Salpeter (BS) fermion-antifermion scattering 
kernels the $\sigma$ and $\pi$ composites can be incorporated
in a RG invariant manner by making use 
of the skeleton expansion, {\it e.g.}, see Ref.~\cite{bjodre}).
Analogous to QED kernels we define the 
one-boson irreducible kernel $K^{(1)}$, and the two-fermion one-boson 
irreducible BS kernel $K^{(2)}$, where these kernels now also include the 
$\sigma$ and 
$\pi$ composites.
For both type of kernels a skeleton expansion exists.
The integral equation between $K^{(1)}$ 
and $K^{(2)}$ is known as the Bethe-Salpeter equation.

\begin{figure}[b!]
\epsfxsize=10cm
\epsffile[-20 300 350 540]{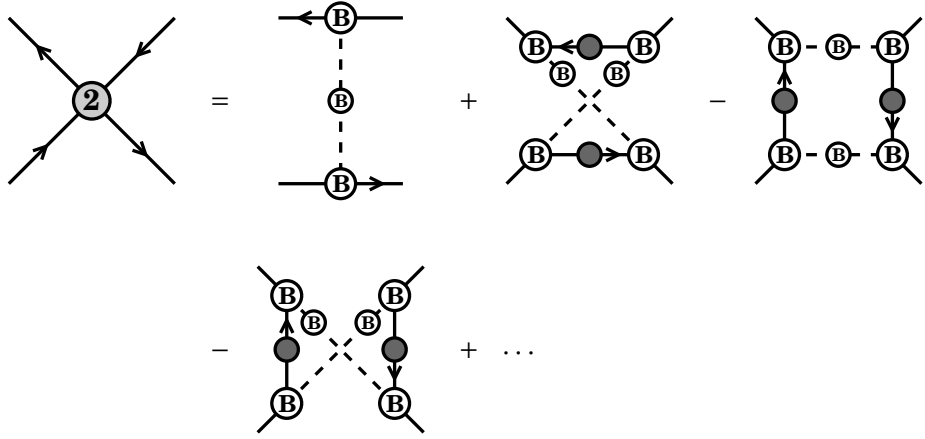}
\caption{Skeleton expansion for the BS kernel $K^{(2)}$.}
\label{fig_skel2}
\end{figure}

The skeleton expansion is a series in topologically distinct Feynman diagrams
with all vertices and propagators fully dressed.
The skeleton expansion is a special way of resumming the entire set of 
Feynman diagrams in a consistent manner, {\em i.e.}, without double counting.
The lowest order terms (``lowest'' in terms of loops) 
of the skeleton expansion for $K^{(2)}$ is 
illustrated in Fig.~\ref{fig_skel2}.
The blobs with the letter ``B'' in the full vertices and propagators 
represent photons, and composite $\sigma$ and $\pi$ exchanges.

Each term in the skeleton expansion of the BS kernel $K^{(2)}$ is RG 
invariant, up to fermion wave function factors, {\em i.e.} the expansion is 
independent of the renormalization factors $Z_3$
and $Z=Z_\sigma=Z_\pi$ (see Eqs.~(\ref{Z1Z2Z3def}) 
and (\ref{Zsigmapidef})) of, respectively, the gauge field and the 
composite fields $\sigma$ and $\pi$.
The two $Z^{-1}$ factors with anomalous dimensions of each Yukawa 
vertex cancel with the $Z^2$ factors of the $\sigma$ and $\pi$ propagators, 
leading to cutoff independent fermion-antifermion scattering amplitudes, 
{\em e.g.}, see Eq.~(\ref{longnatYF}).
\subsection{The {\boldmath$U_L(N)\times U_R(N)$} chiral symmetry and 
the {\boldmath$1/N$} expansion}\label{ssec_cs1/N}
As was mentioned in Sec.~\ref{sec_AGNJL}, the Abelian GNJL model, with 
$N$ number of fermion flavors, is taken 
to be invariant under global $U_L(N)\times U_R(N)$ chiral transformations,
so that both the scalar and pseudoscalar four-fermion interactions are in 
the adjoint representation, and, consequently, 
the number of scalar composites ($N^2$) equals the number of pseudoscalar 
composites ($N^2$). 
In this way, when $N$ is large we can use the $1/N$ expansion introduced 
by 't Hooft \cite{tho74}.
This provides us with a scheme to incorporate four-fermion 
interactions beyond the mean-field approach.
The $1/N$ expansion states that the planar 
({\em i.e.}, ladder) diagrams, with fermions at the edges, 
describe the leading or dominant contributions to Green functions.

The interesting feature of such a $1/N$ expansion is that
Feynman diagrams can be classified in terms of two-dimensional surfaces
with specific topology.
Diagrams with other (than planar) topological structures are suppressed
by at least factors of $1/N$, 
and in the limit of large $N$, their contribution
can be neglected with respect to planar graphs.
One important rule is to draw Feynman graphs with fermion loops 
forming the boundary of the graph (if possible).
In this way, vertex corrections are not necessarily classified as being 
planar.

In the context of the 't Hooft's $1/N$ expansion, 
we should consider internal or virtual $\sigma$ and $\pi$ exchanges 
analogous to the gluon exchanges
with the important difference that due to the chiral symmetry we have two
types of particles both being in the adjoint representation 
($N^2$ scalars and $N^2$ pseudoscalars).
Then, by keeping track of the flavor indices within a particular Feynman 
diagram, we can count factors of $1/N$.
Each fermion carries a flavor index $(i)$, which runs from $1$ to $N$.
A virtual $\sigma$, $\pi$ exchange, being associated with two Yukawa 
vertices, carries two flavor indices.
Therefore, as a result of the Fierz identity (\ref{Fierz}), 
each virtual $\sigma$, $\pi$ exchange  
gives rise to a pair of Kronecker $\delta$ functions
connecting the flavor indices of the scattered fermions.
In the context of flavor indices,
either a $\sigma$ or $\pi$ boson can be considered as a propagating
fermion-antifermion pair carrying double flavor indices. 

Whenever a trace over a flavor Kronecker $\delta$ function enters
into the expression for a particular Feynman diagram,
we speak of an index loop.
An index loop is easily identified by using the double-line representation 
of 't Hooft.
A fermion propagator is represented by a single index line
({\em i.e.}, fermion line), whereas each internal scalar, respectively, 
pseudoscalar propagator is represented by a double index line. 
Consequently, whenever, after drawing a particular Feynman diagram,  
an index line closes, it forms an index loop giving rise to a factor 
$N={\rm Tr}\, \delta$.

The topology of the Feynman diagram determines the factors of $N$.
The vacuum polarization has the topology of a sphere with a single hole 
({\em i.e.}, a disk), 
where the fermion-loop forms the boundary ({\em i.e.}, hole) of the graph.
It can be shown straightforwardly,
that planar diagrams in the vacuum polarization
with $n$ exchanges of $\sigma$'s and $\pi$'s
are associated with a factor $N^{n+1} g_Y^{2n}$, where $g_Y$ is
an ``effective'' Yukawa coupling describing
the interaction of scalars and pseudoscalar with fermions.
For the time being we leave unspecified such a coupling.

In absence of bare mass, the $U_L(N)\times U_R(N)$ symmetry
allows us to write each full Yukawa vertex, photon-fermion vertex, 
fermion propagator, and $\sigma$, $\pi$ boson propagator as
\begin{eqnarray}
&&\Gamma^\alpha_{{S}^{ab}_{ij}}(k,p)=\tau^\alpha_{ij} \Gamma_{{S}ab}(k,p),
\qquad \Delta_{S}^{(\alpha)}(q)=\Delta_{S}(q),\\
&&\Gamma^\alpha_{{P}^{ab}_{ij}}(k,p)=\tau^\alpha_{ij} \Gamma_{{P}ab}(k,p),
\qquad \Delta_{P}^{(\alpha)}(q)=\Delta_{P}(q),\\
&&\Gamma^\mu_{^{ab}_{ij}}(k,p)=\delta_{ij} \Gamma^\mu_{ab}(k,p),
\qquad S^{(i)}(p)=S(p),
\end{eqnarray}
with $a,b$ spinor indices, $i,j$ flavor indices, and $\alpha$ the
$U(N)$-generator 
index (see Refs.~\cite{manuel,gure98} for the definitions of the proper 
vertices and connected two-point Green functions).
So that
\begin{eqnarray}
\sum_{\alpha=0}^{N^2-1}
\Gamma^\alpha_{{S}^{ab}_{ij}}(k+q,k) \Delta_{S}^{(\alpha)}(q)
\Gamma^\alpha_{{S}^{cd}_{kl}}(p,p+q)= \delta_{il} \delta_{kj}
\Gamma_{{S}ab}(k+q,k) \Delta_{S}(q)\Gamma_{{S}cd}(p,p+q),\label{fierzap1}
\end{eqnarray}
because of the Fierz identity (\ref{Fierz}).
Then the first term of the skeleton expansion for $K^{(2)}$ is the following 
single boson exchange term:
\begin{eqnarray}
(-ie_0^2)K^{(2)}_{^{ab,cd}_{i_1j_1,i_2j_2}}(k,p,p+q)
&=&\delta_{i_1j_1}\delta_{i_2j_2}
(-i)\Gamma_{{S}cb}(p+q,p)i\Delta_{S}(q)
(-i)\Gamma_{{S}ad}(k,k+q)\nonumber\\
&+&
\delta_{i_1j_1}\delta_{i_2j_2}
(-i)\Gamma_{{P}cb}(p+q,p)
i\Delta_{P}(q)
(-i)\Gamma_{{P}ad}(k,k+q)\nonumber\\
&+&\delta_{i_2j_1}\delta_{i_1j_2}
(-ie_0)\Gamma^\lambda_{cb}(p+q,p)iD_{\lambda\sigma}(q)
(-ie_0)\Gamma^\sigma_{ad}(k,k+q).\label{BSapprox}
\end{eqnarray} 

As a result of the chiral symmetry the contributions of
four-fermion interactions, which are represented by $\sigma$ and $\pi$ 
exchanges, exhibit two distinct features depending on whether they are 
incorporated in SDE's describing quantities connected with so-called zero-spin 
structures\footnote{Such structures are characterized by spinor matrices 
which commute with the ${\gamma_5}$ matrix.} 
({\em e.g.}, the dynamical mass $\Sigma$, the Yukawa vertices 
$\Gamma_{S}$, $\Gamma_{P}$, and the $\sigma$ and $\pi$ propagators 
$\Delta_{S}$, $\Delta_{P}$),
or whether the exchanges are included in
SDE's describing nonzero-spin structures (anti-commuting with ${\gamma_5}$)
({\em e.g.}, the vacuum polarization $\Pi$, 
the photon-fermion vertex $\Gamma^\mu$, and the fermion wave function 
${\cal Z}=Z_2$). Henceforth, we refer to (non)zero-spin functions, and their 
equations as (non)zero-spin channels. 

The chiral symmetry gives rise to the following properties.
\begin{enumerate}
\item{
In spin-zero-channels, the contribution of 
planar diagrams ({\em i.e.}, planar in $\sigma$ and $\pi$ exchanges)  
vanishes due to the fact
that the exchange of a $\sigma$ has an opposite sign with respect 
to a $\pi$ exchange.
Why?
Let us consider a planar contribution  
to the scalar vacuum polarization which contains (amongst others)
a $\pi$ exchange.
Both ${\gamma_5}$ matrices corresponding to this particular planar $\pi$
exchange can be eliminated from the fermion trace of the scalar vacuum 
polarization by moving them to the right-hand side of the trace. 
For planar diagrams such a process involves the interchange of
the ${\gamma_5}$ matrix with an even number of fermion propagators, and
an arbitrary number of Yukawa vertices.
Since the Yukawa vertices commute with the ${\gamma_5}$ matrix,
and ${\gamma_5}$ anti-commutes with the 
fermion propagator\footnote{In the symmetric phase 
${\gamma_5} S=-S{\gamma_5}$.} $S$,
the process of moving the ${\gamma_5}$ to the right does not introduce an 
overall minus sign.
Now using that 
$(i{\gamma_5})(i{\gamma_5})=-(\bf 1)(\bf 1)$, we see that the diagram 
containing a specific planar $\pi$ exchange
is identical to minus the same diagram with the $\pi$ exchange replaced by 
a $\sigma$ exchange.
Since each diagram containing a $\pi$ exchange has a scalar counter part
({\em i.e.}, an analogous diagram with a $\sigma$ 
instead of a $\pi$ exchange), the sum of all planar diagrams, 
with a particular number of exchanges, vanishes.
}
\item{
In nonzero-spin channels (think of $\Pi$, $\Gamma^\mu$, etc.)
containing vertices which anti-commute with the ${\gamma_5}$ matrix, 
the situation is different: planar $\sigma$ and $\pi$ exchanges contribute
with identical sign.
Let us now consider a planar contribution  
to the (photon) vacuum polarization containing 
a $\pi$ exchange.
If we again move the ${\gamma_5}$ matrices to the right-hand side of the
trace, we get an overall minus sign due to the anti-commutation
of ${\gamma_5}$ with $\gamma^\mu$.
This means that any planar diagram in the vacuum polarization
containing a $\pi$ exchange is identical to the same diagram with 
the $\pi$ exchange replaced by a $\sigma$ exchange.
}
\end{enumerate}
The properties described above are, strictly speaking, only valid in 
the symmetric (massless) phase, where 
the ($2N^2$) $\sigma$ and $\pi$ bosons are degenerate.
However, in the broken phase, the properties are valid whenever
momenta larger than the dynamical mass $\Sigma$ or $m_\sigma$ are considered, 
because then the degeneracy emerges too.

These properties also provide us with a general argument
why the mean-field approach for four-fermion operators
for Green functions corresponding to spin-zero channels 
({\em e.g.}, ${\Gamma_{S}}$ and $\Delta_{S}$)
is reliable. For such channels planar contributions vanish
and the next non-vanishing contributions (such as contributions
containing crossed $\sigma$ and $\pi$ exchanges)
are proportional to $1/N$, thus small for large $N$.
This implies that quantities such as the critical curve, dynamical mass, 
anomalous dimensions etc., are nearly independent of $N$, and are described 
rather well by the mean-field approach.
To the contrary, the cancellation of scalars against pseudoscalars degrees of 
freedom does not occur in the vacuum 
polarization $\Pi$ which is a nonzero-spin channel.
\subsection{The fermion wave function}
The inclusion of relevant four-fermion interactions beyond
the mean-field approach requires a reinvestigation of the SDE for the 
fermion wave function ${\cal Z}=Z_2$ (Eq.~(\ref{Z1Z2Z3def})),
\begin{eqnarray}
S(p)=\frac{{\cal Z}(p^2)}{{\hat p}-\Sigma(p^2)},
\end{eqnarray}
with $S$ the fermion propagator and $\Sigma$ the dynamical mass.
In QED in the quenched-ladder approximation, the fermion wave function has a 
gauge dependent anomalous dimension. In the Landau gauge, this anomalous 
dimension vanishes and ${\cal Z}=1$. 

We conjecture that the inclusion of relevant four-fermion interactions
does not introduce an anomalous dimension for the fermion wave function other
than already introduced by the gauge interactions.
Thus, in the Landau gauge, the wave function ${\cal Z}$ is finite
though it might deviate from unity.
The main argument in support of the conjecture stated above is that
only one full Yukawa vertex appears in the self-energy part, which means 
that anomalous dimensions of four-fermion interactions are not canceled.
Only two fully dressed Yukawa vertices and a fully dressed scalar composite 
are RG invariant (anomalous dimensions cancel).
Consequently a remnant power of the cutoff (related to anomalous 
dimension of a Yukawa vertex) lowers the degree of divergence of the 
self-energy part from a logarithmic divergence to a finite integral.
Therefore, throughout this paper we assume ${\cal Z}=1$.

A nice feature of the assumption that ${\cal Z}=1$ is that
with the gauge interaction treated in the quenched-ladder approach
the chiral and vector Ward--Takahashi identities (WTI's) are preserved, 
since in channels with spin-zero the planar $\sigma$ and $\pi$ exchanges 
cancel each other.
\section{Scalars, pseudoscalars, and charge screening}\label{sec_gHY}
Since, in the GNJL model, the scalars and pseudoscalars are neutral states 
which therefore do not couple to the photon field, 
their contribution to the vacuum polarization is described indirectly 
in terms of photon-fermion vertex corrections, and fermion self-energy 
corrections.
Hence, in order to gain some intuition for the role of scalar degrees of 
freedom on the mechanism of charge screening, 
we analyze the two-loop contribution arising from $\sigma$ and $\pi$ exchanges
to the vacuum polarization. 

\begin{figure}[t!]
\epsfxsize=10cm
\epsffile[0 300 370 540]{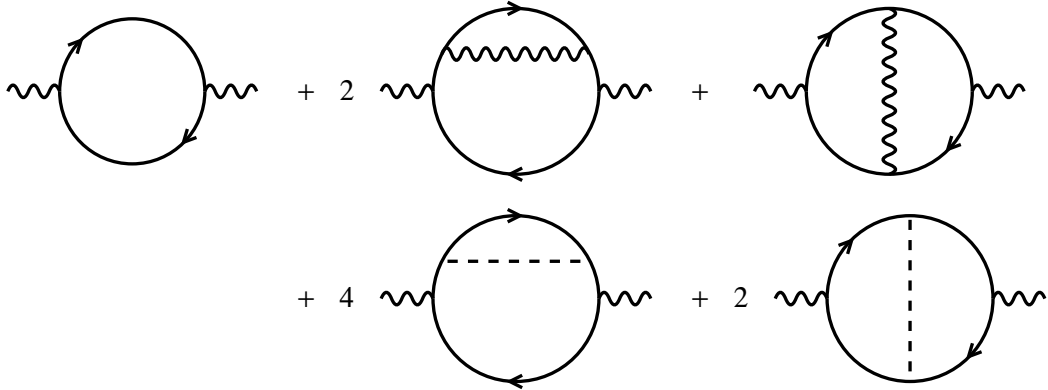}
\caption{
One and two-loop vacuum polarization corrections containing both
an internal photon (wavy) and a $\sigma$ and $\pi$ exchange (dashed).
The contribution of $\pi$ 
equals the contribution of $\sigma$.}
\label{fig_twoloop}
\end{figure}

Let us consider an Abelian 
gauge--Higgs--Yukawa type of interaction 
described by the Lagrangian\footnote{For 
a discussion of the renormalizability of 
non-Abelian gauge--Higgs--Yukawa models and non-Abelian GNJL models
we refer to Refs.~\cite{hakikuna94}, and references therein.}
\begin{eqnarray}
{\cal L}_{GHY}&=&-\frac{1}{4}F_{\mu\nu}F^{\mu\nu}
+\bar\psi i\gamma^\mu \partial_\mu\psi+\frac{1}{2}(\partial_\mu\sigma)^2
+\frac{1}{2}(\partial_\mu\pi)^2\nonumber\\
&-&e_0\bar\psi \gamma^\mu A_\mu \psi-g_Y\bar\psi(\sigma+i{\gamma_5}\pi)\psi
-V(\sigma,\pi), \label{GHY}
\end{eqnarray}
where the potential $V$ contains, {\em e.g.}, mass terms, and a $\sigma^4$ type 
of interaction ({\em i.e.} a quartic scalar interaction).
For simplicity,  
we ignore the effect of the potential $V$.
In Appendix~\ref{appvacpol}, the two-loop contribution to $\Pi$
has been computed for the special case of $N=1$, see also 
Fig.~\ref{fig_twoloop}.
If the scalar and pseudoscalar fields in Eq.~(\ref{GHY}) 
are both in the adjoint representation of $U(N)$, 
the result, for arbitrary $N$, reads
\begin{eqnarray}
\Pi\left[q/\Lambda,\alpha_0,\lambda_Y\right]&\approx& \frac{N\alpha_0}{\pi}
\left(\frac{2}{3}+\frac{\alpha_0}{2\pi}-\frac{N\lambda_Y}{2\pi}\right)
\ln \frac{\Lambda}{q}
+(\alpha_0/\pi){\cal O}(1),\label{2loopwscalarN}
\end{eqnarray}
with $\lambda_Y=g_Y^2/4\pi$.
The $\beta$ function corresponding to such a vacuum polarization 
(Fig.~\ref{fig_twoloop}) is
\begin{eqnarray}
\beta_\alpha(\alpha_0,\lambda_Y)=\frac{N\alpha_0^2}{\pi}
\left(\frac{2}{3}+\frac{\alpha_0}{2\pi}-\frac{N\lambda_Y}{2\pi}\right).
\label{betscal}
\end{eqnarray}
The interesting result of this computation is difference in sign
between terms corresponding to photon exchanges, and terms corresponding 
to (pseudo)scalar exchanges. 
From this point of view,
the fundamental scalars and pseudoscalars in a gauge-Higgs-Yukawa system
tend to decrease charge screening.
Furthermore, we might be tempted to conclude that a nontrivial
root of Eq.~(\ref{betscal}) could be realized whenever 
$N\lambda_Y/2\pi\sim 2/3$.
However, the complete situation is more involved. 
The RG equation for, {\em e.g.}, $\lambda_Y$ should be
considered too, {\em i.e.}, we should compute the $\beta$ functions
of $\lambda_Y$, and of any quartic scalar coupling.
If and only if a nontrivial (nonzero) UV stable fixed point for 
$\lambda_Y$ exists, 
the realization of a zero of Eq.~(\ref{betscal}) becomes a realistic option. 
In other words, such a scenario is only possible if the Yukawa interaction 
$\lambda_Y$ is nontrivial.
The discussion in Sec.~\ref{sec_ren_hyp} 
implies that in order to obtain a nontrivial Yukawa coupling, the 
hyperscaling laws should be obeyed.
The idea is that, 
nonperturbatively, close to the critical curve in 
the GNJL model, 
the scalar and pseudoscalar Yukawa interactions are nontrivial, 
and kinetic terms for 
the scalar and pseudoscalar composites are effectively induced
via the appearance of a large anomalous dimension.
\section{The vacuum polarization in the {\boldmath$1/N$} 
expansion}\label{vacpol1n}
The purpose of the present paper is to investigate the existence of an 
UV stable fixed point $(\alpha_\star,g_\star)$ of the gauge coupling, so that
\begin{eqnarray}
\beta_\alpha(\alpha_\star,g_\star)&=&0,\qquad \eta_\alpha\equiv 
-\beta^\prime_\alpha(\alpha_\star,g_\star)>0, \label{UVfixdef2}
\end{eqnarray}
with $\eta_\alpha$ a critical index characterizing the RG flow in the 
neighborhood of $(\alpha_\star,g_\star)$.
With $\alpha$ close to $\alpha_\star$ the $\beta$ function 
linearizes as 
\begin{eqnarray}
\beta_\alpha(\alpha,g_\star)\approx\eta_\alpha (\alpha_\star-\alpha).
\label{betlin}
\end{eqnarray}
The $\beta$ function for the gauge coupling follows from the RG transformation
relating the charge $\alpha_x$ at scale $x$ to the charge $\alpha_y$ 
at scale $y$ via
\begin{eqnarray}
\alpha_x={\cal R}_\alpha\left[x/y,\alpha_y,g_y\right]=\frac{\alpha_y}{1
+\Pi\left[x/y,\alpha_y,g_y\right]},
\qquad \Pi\left[1,\alpha_x,g_x\right]=0, \label{RGdef1}
\end{eqnarray}
where $\Pi(x^2)=\Pi\left[x/y,\alpha_y,g_y\right]$ is the (Euclidean)
vacuum polarization.
The RG transformation should satisfy the RG semigroup property
(with $x<y<z$)
\begin{eqnarray}
{\cal R}_\alpha\left[x/y,\alpha_y,g_y\right]&=&
{\cal R}_\alpha\left[x/z,\alpha_z,g_z\right]=
{\cal R}_\alpha\left[x/y,{\cal R}_\alpha\left[y/z,\alpha_z,g_z\right],
{\cal R}_g\left[y/z,\alpha_z,g_z\right]\right], \label{RGsemi1}\\
{\cal R}_\alpha\left[x/y,\alpha_\star,g_\star\right]&=&
{\cal R}_\alpha\left[x/z,\alpha_\star,g_\star\right],\label{RGsemi2}
\end{eqnarray}
where ${\cal R}_g$ is the RG transformation for $g$, satisfying 
analogous equations.
Then the $\beta$ function for $\alpha$ is defined as
\begin{equation}
\beta_\alpha(\alpha_x,g_x)\equiv\frac{
\partial{\cal R}_\alpha\left[w,\alpha_x,g_x\right]}{
\partial \ln w}\Biggr|_{w=1}.
\end{equation}
The RG semigroup property (\ref{RGsemi1}) gives rise to the well known 
differential RG equation 
\begin{eqnarray}
\beta_\alpha(\alpha_y,g_y)=y\frac{d\alpha_y}{dy}, \label{RGdiffeq2}
\end{eqnarray}
whose solution is
\begin{eqnarray} 
\alpha_x=\sum_{n=0}^\infty\frac{1}{n!}\left(\ln \frac{x}{y}\right)^n
\left[\beta_\alpha(\alpha_y,g_y)\frac{\partial }{
\partial\alpha_y}\right]^n\alpha_y. \label{RGsol1}
\end{eqnarray} 
To obtain a nontrivial theory in the IR the existence of an UV stable 
fixed point is required. 
Close to the UV stable fixed point ($\beta_\alpha=\beta_g=0$),
with $g_y=g_\star$, we have that
\begin{eqnarray} 
\left[\beta_\alpha(\alpha_y,g_\star)
\frac{\partial}{\partial\alpha_y}\right]^n\alpha_y
\approx \beta_\alpha(\alpha_y,g_\star)
\left(\beta^\prime_\alpha(\alpha_\star,g_\star) \right)^{n-1}+
{\cal O}\left(\beta^2\right),
\end{eqnarray} 
thus
\begin{eqnarray}
\alpha_x\approx \alpha_\star
+(\alpha_y-\alpha_\star)\left(\frac{y}{x}\right)^{\eta_\alpha},
\label{sclfixpa}
\end{eqnarray}
which is the solution of Eq.~(\ref{RGdiffeq2}) 
using Eq.~(\ref{betlin}).
The above expressions are only valid if
both $\alpha_x$ and $\alpha_y$ are 
in the neighborhood of the UV stable fixed point $\alpha_\star$, 
therefore we have the fine-tuning condition:
\begin{eqnarray}
\Biggr|\frac{(\alpha_y-\alpha_\star)}{\alpha_\star}
\left(\frac{y}{x}\right)^{\eta_\alpha}\Biggr|\ll 1,\label{ftineq}
\end{eqnarray}
which in case of $y\gg x$ implies 
that $\alpha_y$ is tuned sufficiently close to $\alpha_\star$.

In case of the Abelian GNJL model the gauge coupling $\alpha_q$
in the infrared (IR) region is related to the bare gauge coupling 
$\alpha_0$ via the RG transformation 
$\alpha_q={\cal R}_{\alpha}\left[q/\Lambda,\alpha_0,g_0\right]$,
where the four-fermion interactions contribute to the vacuum polarization 
$\Pi$ (Fig.~\ref{fig_vacpol}) through the full photon fermion vertex, see 
Fig.~\ref{fig_vertex}.
Assuming that the hyperscaling laws are satisfied,
we can write the vacuum polarization for 
$\Lambda\gg q\geq |m_\sigma|$ ({\em i.e.}, near $g_c(\alpha_0)$) as
\begin{eqnarray}
\Pi\left[q/\Lambda,\alpha_0,g_0\right]=
f_{(1)}(\alpha_0,g_0) \ln\frac{\Lambda}{q}+
f_{(2)}(\alpha_0,g_0) \left(\ln\frac{\Lambda}{q}\right)^2
+f_{(3)}(\alpha_0,g_0) \left(\ln\frac{\Lambda}{q}\right)^3+\dots,
\label{genvacpolexpr}
\end{eqnarray}
where each factor of $\ln \Lambda/q$ corresponds to 
a single fermion loop with two outgoing photon lines.
Thus the function $f_{(1)}$ represents the contribution of all 
diagrams to $\Pi$ in which the internal photon propagators are replaced 
by the bare or canonical form.

In order for $\Pi$ to give rise to a RG transformation satisfying 
Eqs.~(\ref{RGdef1}) and (\ref{RGsemi1}),
it can be derived from Eq.~(\ref{RGsol1}) that
the functions $f_{(1)}$, $f_{(2)}$, and $f_{(3)}$ 
should be related to $\beta_\alpha$ in the following way:
\begin{eqnarray} 
\alpha f_{(1)}(\alpha,g)&=&\beta_\alpha(\alpha,g),\label{beaf} \\
f_{(1)}^2-f_{(2)}&=&\frac{1}{2\alpha} \beta_\alpha \beta_\alpha^\prime,
\label{f2}\\
f_{(1)}^3-2f_{(1)} f_{(2)}+f_{(3)}
&=&\frac{1}{6\alpha} 
\left[\beta_\alpha\frac{\partial}{\partial \alpha}\right]^2\beta_\alpha,\quad 
\dots. \label{f3etc}
\end{eqnarray} 
These identities are nontrivial and require a high degree 
of self-consistency of the theory in the form of Ward 
identities.\footnote{The proof of Eqs.~(\ref{beaf})--(\ref{f3etc}) can 
be performed order by order within perturbation theory.}
Since, within our approximation scheme,
the Ward identities are respected, we assume that 
Eqs.~(\ref{beaf})--(\ref{f3etc}) are satisfied.
Equation~(\ref{beaf}) now relates the $\beta$ function to the
function $f_{(1)}$. Clearly
the $\beta$ function has a Gaussian or trivial fixed point 
at $\alpha_0=0$.

From this we also have that with $\alpha_0\rightarrow \alpha_\star$
at $g_0=g_\star$ (see Eqs.~(\ref{betlin}), (\ref{sclfixpa}), 
and (\ref{ftineq})):
\begin{eqnarray}
\Pi\left[q/\Lambda,\alpha_0,g_\star\right]
\approx \frac{\beta_\alpha(\alpha_0,g_\star)}{\alpha_\star \eta_\alpha}
\left[-1+\left(\frac{\Lambda}{q}\right)^{\eta_\alpha}\right],\qquad
\frac{\beta_\alpha(\alpha_0,g_\star)}{\alpha_\star \eta_\alpha}
\left[-1+\left(\frac{\Lambda}{q}\right)^{\eta_\alpha}\right]\ll 1.
\label{quenchhyp2}
\end{eqnarray}
Hence, $\Delta(q)$ of Eq.~(\ref{bargb}) is
\begin{eqnarray}
\Delta(q)\approx
\frac{1}{q^2}
\left\{1-\frac{\beta_\alpha(\alpha_0,g_\star)}{\alpha_\star \eta_\alpha}
\left[-1+\left(\frac{\Lambda}{q}\right)^{\eta_\alpha}\right]\right\}.
\label{photasymuv}
\end{eqnarray}
The second term on the right-hand side of Eq.~(\ref{photasymuv})
will only contribute via internal photon propagators
to the functions $f_{(2)}$, $f_{(3)}$, etc. and not to the 
function $f_{(1)}$.
In order to find an UV stable fixed point of $\beta_\alpha$, we only have to 
compute $f_{(1)}$ and, therefore, we neglect all corrections to $\Delta(q)$ 
other than canonical in internal photon propagators.
Because the contribution of $\Pi$ is neglected, 
this procedure is identical to quenching  
the internal photon propagators, although
it is not quenched 
in the sense of taking $N\rightarrow 0$!

Now in correspondence with Eq.~(\ref{betaa0gc})  
we search for UV stable roots of Eq.~(\ref{beaf}), {\em i.e.},
$f_{(1)}(\alpha_0,g_c(\alpha_0))=0$.
In case of pure QED (without four-fermion coupling) 
the function $f_{(1)}$ has been studied thoroughly by Johnson {\em et al.} in
Refs.~\cite{jowiba67,bajo69,joba73} and by Adler \cite{ad72} in the 
context of massless QED.  
In Ref.~\cite{jowiba67}, an expression for $f_{(1)}$ is obtained 
in term of the BS kernel $K^{(2)}$ (Sec.~\ref{sec_beyondqmf}) 
as the single unknown Green function.

We mention that, although the strong belief of 
Johnson {\em et al.} in the possible existence of finite QED 
seems poorly motivated from the point of view 
of Wilson's RG methods \cite{wiko74},\footnote{
The authors of \cite{jowiba67} do not address 
the dynamical origin of the singular critical behavior
({\em e.g.}, D$\chi$SB), 
which would be required for the realization of an UV stable fixed point 
in QED.} their methods and techniques are sound and 
directly applicable to the GNJL model.

In Appendix~\ref{dervJWB}, we expose a brief derivation
of the result of Ref.~\cite{jowiba67} and point out the applicability 
to the Abelian GNJL model.  
Then, by taking into account also relevant four-fermion interactions at 
$g_0=g_c(\alpha_0)$ via the BS kernel, we can derive 
from Eqs.~(\ref{fdef1})
and (\ref{beaf}) that 
\begin{eqnarray}
\beta_\alpha(\alpha_0,g_c)=\frac{N\alpha_0^2}{\pi}
\left[\frac{2}{3}+
\frac{\phi_1+\phi_2(2+\phi_2)}{1-\phi_1}+\phi_3
\right],\label{betasc2a}
\end{eqnarray}
where the functions $\phi_1$, $\phi_2$, and $\phi_3$ are 
defined as follows:
\begin{eqnarray}
\phi_1&\equiv& -\lim_{\Lambda\rightarrow \infty}\frac{ie_0^2}{48N}
\int_{\Lambda}\frac{{d}^4p}{(2\pi)^4}\,
{\rm \tilde Tr}\left[
\frac{(\gamma^\mu{\hat p} \gamma^\alpha
-\gamma^\alpha{\hat p}\gamma^\mu)}{2p^4}
K^{(2)}(p,k)(\gamma_\mu{\hat k} \gamma_\alpha-\gamma_\alpha{\hat k}\gamma_\mu)
\right],\label{phi1}
\\
\phi_2&\equiv& -
\lim_{\Lambda\rightarrow \infty}
\frac{ie_0^2}{48N}
\int_\Lambda\frac{{d}^4p}{(2\pi)^4}\,
{\rm \tilde Tr}\left[
\frac{{\hat p}\gamma^\mu {\hat p}}{p^4}
K^{(2)\alpha}(p,k)(\gamma_\mu{\hat k} \gamma_\alpha
-\gamma_\alpha{\hat k}\gamma_\mu)\right],\label{phi2}\\
\phi_3&\equiv&\lim_{\Lambda\rightarrow \infty} 
\frac{ie_0^2}{48N}\int_\Lambda\frac{{d}^4p}{(2\pi)^4}\,
{\rm \tilde Tr}\left[
\frac{{\hat p}\gamma^\mu {\hat p}}{p^4}K_\alpha^{(2)\alpha}(p,k) 
{\hat k}\gamma_\mu{\hat k}\right],\label{phi3}
\end{eqnarray}
with
\begin{eqnarray}
K^{(2)}(p,k)&\equiv& K^{(2)}(p,p+q,k+q)\biggr|_{q=0},\label{k0derdef}\\
K^{(2)\alpha}(p,k)
&\equiv& \frac{\partial}{\partial q_\alpha} 
K^{(2)}(p,p+q,k+q)\biggr|_{q=0},\label{k1derdef}
\\
K_\alpha^{(2)\alpha}(p,k)
&\equiv& 
\frac{\partial^2}{\partial q^\alpha\partial q_\alpha}
K^{(2)}(p,p+q,k+q)\biggr|_{q=0}.\label{k2derdef}
\end{eqnarray}
The trace over spinor and flavor indices is defined as
\begin{eqnarray}
{\rm \tilde Tr}\left[L(p)K(p,k)R(k)\right]\equiv  
L_{dc}(p) K_{^{cd,ab}_{ii,jj}}(p,k)R_{ba}(k),
\end{eqnarray}
with $L$ and $R$ some projectors and with appropriate summation
over double spinor (4) and flavor indices (N).

The BS kernels in Eqs.~(\ref{phi1})--(\ref{phi3}) 
contain, in principle, all diagrams except those corresponding
to vacuum polarization corrections, since all internal photon propagators 
are canonical or ``quenched.''
As was pointed out in Sec.~\ref{sec_beyondqmf},
the $1/N$ expansion states that the planar diagrams for
the $\sigma$ and $\pi$ exchanges are dominant.
The approximation for the BS kernel $K^{(2)}$, which
generates the entire set of planar scalar and pseudoscalar skeleton 
diagrams including ladder photon exchanges 
for the vacuum polarization is the following:
the BS kernel $K^{(2)}$ is approximated by its ``lowest'' order 
skeleton graph, {\em i.e.}, 
\begin{eqnarray}
K^{(2)}_{^{cd,ab}_{kl,ij}}(p,p+q,k+q)
&=&\frac{\delta_{ij}\delta_{kl}}{e_0^2}
\biggr[\Gamma_{{S}cb}(k+q,p+q) \Delta_{S}(k-p) \Gamma_{{S}ad}(p,k)\nonumber\\
&+&\Gamma_{{P}cb}(k+q,p+q) \Delta_{P}(k-p) \Gamma_{{P}ad}(p,k)\biggr]
\nonumber\\&+& 
\delta_{il}\delta_{kj}  \gamma^\mu_{ad}\gamma^\nu_{cb} D_{\mu\nu}(k-p).
\end{eqnarray}

In the symmetric phase, the decomposition of the scalar Yukawa vertex 
${\Gamma_{S}}$ is given by Eq.~(\ref{vertfiesdef}).
Furthermore, due to chiral symmetry, we have
the identities
\begin{eqnarray}
(i{\gamma_5}){\Gamma_{S}}(k,p)={\Gamma_{P}}(k,p),\qquad 
\Delta_{S}(q)=\Delta_{P}(q),
\end{eqnarray}
the $\sigma$ and $\pi$ propagators are degenerate.
Second, it was shown in Ref.~\cite{gure98} that the structure function 
$F_2$ is rather small compared to the leading structure function $F_1$
(it is assumed that $F_1$  describes the leading asymptotic behavior of 
the Yukawa vertices).
Therefore, we neglect contributions related to the scalar structure function 
$F_2$.
Although it might be possible 
that the contribution coming from gauge interactions is smaller 
than corrections resulting from this structure function $F_2$, we 
keep the gauge interaction in order to compare with
results mentioned in the literature.
Thus, we take for $K^{(2)}$
\begin{eqnarray}
K^{(2)}_{^{cd,ab}_{kl,ij}}(p,p+q,k+q)
&\approx&\frac{\delta_{ij}\delta_{kl}}{e_0^2}
F_1(k+q,p+q)F_1(p,k)\Delta_{S}(k-p)\left[{\bf 1}_{ad}{\bf 1}_{cb}
+{i{\gamma_5}}_{ad}{i{\gamma_5}}_{cb}\right]\nonumber\\
&+&\delta_{il}\delta_{kj} D_{\mu\nu}(k-p)\gamma^\mu_{ad}\gamma^\nu_{cb}.
\label{K2approx}
\end{eqnarray}

With this truncation for the BS kernel $K^{(2)}$, we can actually 
compute the $\phi_j$ functions (\ref{phi1})--(\ref{phi3}), and subsequently
analyze the $\beta$ function (\ref{betasc2a}).
The truncation (\ref{K2approx}) generates an infinite
series of planar contributions to the vacuum polarization
as the leading order in $1/N$, see Fig.~\ref{fig_vacpol1overN}.
As was discussed in Sec.~\ref{ssec_cs1/N} and shown in 
Appendix~\ref{appvacpol} up to two loops, 
the scalars and pseudoscalars give the same contribution in the functions 
$\phi_j$.
The trace over flavor indices yields an overall factor of $N$ in the 
expressions for $\phi_j$ for contributions corresponding 
to $\sigma$ and $\pi$ exchanges.
\section{Computation of the {\boldmath $\beta$} function}\label{sec_compbeta}
In computing the functions $\phi_1$, $\phi_2$, and $\phi_3$,
we initially neglect the ladder photon exchange given in Eq.~(\ref{K2approx}).
Since such contributions were already computed in Ref.~\cite{jowiba67}, it 
will be rather easy to include them later in the analysis.

It is straightforward to show that $\phi_2$ vanishes, within the proposed 
approximations.
Using Eq.~(\ref{K2approx}), we obtain from Eqs.~(\ref{phi2}) and 
(\ref{k1derdef})
that
\begin{eqnarray}
{\rm Tr}\left[{\hat p}\gamma^\mu {\hat p}
K^{(2)\alpha}(p,k)(\gamma_\mu{\hat k} \gamma_\alpha
-\gamma_\alpha{\hat k}\gamma_\mu)\right]
\propto
{\rm Tr}\left[{\hat p}\gamma^\mu {\hat p}
(\gamma_\mu{\hat k} \gamma_\alpha-\gamma_\alpha{\hat k}\gamma_\mu)\right]=0.
\end{eqnarray}
Thus $\phi_2(\alpha_0,g_c)=0$.

With Eq.~(\ref{K2approx}), Eq.~(\ref{phi1}) for $\phi_1$ reads
\begin{eqnarray}
\phi_1(\alpha_0,g_c)&=&-\lim_{\Lambda\rightarrow \infty}
\frac{i}{48N}\int_\Lambda\frac{{d}^4p}{(2\pi)^4}
\frac{1}{2p^4}\sum_{i=1}^N \sum_{j=1}^N
\delta_{ii}\delta_{jj}F_1(k,p)F_1(p,k)\Delta_{S}(k-p)
\nonumber\\
&&
\times \biggr\{ {\rm Tr}\left[
(\gamma^\mu{\hat p} \gamma^\alpha-\gamma^\alpha{\hat p}\gamma^\mu)
(\gamma_\mu{\hat k} \gamma_\alpha
-\gamma_\alpha{\hat k}\gamma_\mu)\right]\nonumber\\
&&+
{\rm Tr}\left[
(\gamma^\mu{\hat p} \gamma^\alpha-\gamma^\alpha{\hat p}\gamma^\mu)
i\gamma_5
(\gamma_\mu{\hat k} \gamma_\alpha
-\gamma_\alpha{\hat k}\gamma_\mu)i\gamma_5\right]
\biggr\}
\nonumber\\
&=&
\lim_{\Lambda\rightarrow \infty}
2Ni\int_\Lambda\frac{{d}^4p}{(2\pi)^4}\frac{(p+k)\cdot k}{(p+k)^4} 
[F_1(k+p,k)]^2\Delta_{S}(p),
\end{eqnarray}
where we have performed a ``harmless''\footnote{The integral is finite, 
therefore translationally invariant.} shift of integration,
and used the fact that $F_1$ is symmetric in the fermion momenta, 
$F_1(p,k)=F_1(k,p)$, because of $C$-$PT$ invariance.
The overall factor $N$ results from tracing the flavor Kronecker 
$\delta$ functions, which is equivalent to closing index lines, 
see Sec.~\ref{ssec_cs1/N}.
After a Wick rotation 
\begin{eqnarray}
\phi_1(\alpha_0,g_c)&=&\lim_{\Lambda\rightarrow \infty}
\frac{N}{8\pi^2}
\int\limits_0^{\Lambda^2}
{d}p^2\,\int \frac{{d}\Omega_p}{2\pi^2}\,
\frac{p^2(p\cdot k+k^2)}{(p+k)^4}[F_1(k+p,k)]^2\Delta_{S}(p).
\end{eqnarray}

Since the integrals for the functions $\phi_j$ are finite,
we can write
\begin{eqnarray}
\phi_1(\alpha_0,g_c)
=-N\int\limits_0^{\infty}{d}u\, G_1(u),
\label{contlimphi1}
\end{eqnarray}
where $u=p^2/k^2$, and
\begin{eqnarray}
G_1(p^2/k^2)\equiv
-\lim_{\Lambda\rightarrow \infty}
\frac{1}{8\pi^2} 
\int \frac{{d}\Omega_p}{2\pi^2}\,
\frac{k^2 p^2(k\cdot p+k^2)}{(k+p)^4}[F_1(k+p,k)]^2\Delta_{S}(p).\label{G1def}
\end{eqnarray}
The function $G_1$ is defined with a minus sign to 
make it a positive function, as will be shown to be the case later.
The angular integral can be performed if we make use of the following 
Chebyshev expansion:
\begin{eqnarray}
\frac{k^2(k\cdot p+k^2)}{(k+p)^4}=\sum_{n=0}^\infty c_n(k^2,p^2) 
U_n(\cos\alpha),\quad\cos\alpha=\frac{k\cdot p}{kp},
\end{eqnarray}
where
\begin{eqnarray}
c_n(k^2,p^2)&=&\frac{2}{\pi}\int\limits_0^\pi{d}\alpha\,
\sin^2\alpha\,U_n(\cos\alpha)\frac{k^2(k\cdot p+k^2)}{(k+p)^4},\\
c_n(k^2,p^2)&=& \frac{(-1)^n}{2}
\left[(2+n)\theta(k^2-p^2)\left(\frac{p}{k}\right)^n
-n\theta(p^2-k^2)\left(\frac{k}{p}\right)^{n+2}\right].\label{cnpr}
\end{eqnarray}
The Chebyshev expansion for the function $F_1$ 
was already introduced in Eq.~(\ref{chebF1}) (and \cite{gure98}).
Thus, following analogous derivations in the Appendix of Ref.~\cite{gure98},
the function $G_1$ can be expressed as
\begin{eqnarray}
G_1(p^2/k^2)=-\lim_{\Lambda\rightarrow \infty}
\frac{p^2\Delta_{S}(p)}{8\pi^2} 
\sum_{l,m,n=0}^\infty C_{lmn} \,c_l(k^2,p^2) f_m(k^2,p^2) f_n(k^2,p^2), 
\label{G1exp}
\end{eqnarray}
where the fully symmetric index 
$C_{lmn}=1$, if $l+m+n=\mbox{even}$ 
and a triangle with sides $l$, $m$, $n$ exists, {\em i.e.},
$|l-m|\leq n \leq l+m$. 
Otherwise $C_{lmn}=0$.

We approximate $G_1$ by keeping only the
lowest order term in the Chebyshev expansion,
\begin{eqnarray}
G_1(p^2/k^2)\approx-\lim_{\Lambda\rightarrow \infty}
\frac{p^2\Delta_{S}(p)}{8\pi^2} 
c_0(k^2,p^2) \left[f_0(k^2,p^2)\right]^2, \label{G1approx}
\end{eqnarray}
where $f_0$ is decomposed into 
the two channel functions $F_{\rm UV}$ and $F_{\rm IR}$, 
see Eq.~(\ref{channelapprox}).
Then, the asymptotics, $k^2\gg p^2$,  respectively, $p^2\gg k^2$,
of $G_1$ are well approximated by the lowest order Chebyshev 
term (\ref{G1approx}).
Again, this is the two channel approximation for the Yukawa vertices 
\cite{gure98}.
However, for momenta $k^2\sim p^2$ the channel approximation 
is not necessarily valid. So, how about $G_1(1)$?
Since, from the appendix of Refs.~\cite{gure98},
it follows that the Chebyshev coefficients
\begin{eqnarray}
f_{2n}(k^2,p^2)\geq 0, \qquad f_{2n+1}(k^2,p^2)\leq 0,
\end{eqnarray}
and from Eq.~(\ref{cnpr}) that
\begin{eqnarray}
c_n(k^2,k^2)=\frac{(-1)^n}{2}\quad\longrightarrow\quad
c_{2n}(k^2,k^2)\geq 0, \quad c_{2n+1}(k^2,k^2)\leq 0.
\end{eqnarray}
Hence, by taking into account the properties of $C_{lmn}$,
we conclude that all terms of the series
\begin{eqnarray}
\sum_{l,m,n=0}^\infty C_{lmn} \,c_l(k^2,k^2) f_m(k^2,k^2) f_n(k^2,k^2)
\end{eqnarray}
of $G_1$ are positive,
and the lowest order term gives a lower bound on the series,
\begin{eqnarray}
c_0(k^2,k^2)\left[f_0(k^2,k^2)\right]^2 \leq
\sum_{l,m,n=0}^\infty C_{lmn} c_l(k^2,k^2) f_m(k^2,k^2) f_n(k^2,k^2).
\end{eqnarray}
Therefore, the approximation Eq.~(\ref{G1approx}) is reliable
for the asymptotics $k^2\gg p^2$, and $p^2\gg k^2$. Moreover, 
Eq.~(\ref{G1approx}) is a lower bound on Eq.~(\ref{G1exp}) at $k^2=p^2$,  
so that at least we will not overestimate the contribution
of scalar and pseudoscalar composites to the vacuum polarization.

The function $G_1$ can now be computed, since $f_0$ is expressed in 
terms of the channel functions $F_{\rm UV}$ and $F_{\rm IR}$
of Eq.~(\ref{channelapprox}).
Furthermore, from Eq.~(\ref{cnpr}) we see that
\begin{eqnarray}
c_0(k^2,p^2)&=& \theta(k^2-p^2),\label{c0}
\end{eqnarray}
and the only nonzero contribution to $G_1$ of Eq.~(\ref{G1approx}) 
comes from the momenta $k^2\geq p^2$. Thus, using Eqs.~(\ref{G1approx}),
(\ref{channelapprox}), and (\ref{c0}), we find
\begin{eqnarray}
G_1(p^2/k^2)&\approx& 
-\lim_{\Lambda\rightarrow \infty}
\frac{p^2\Delta_{S}(p)}{8\pi^2} [F_{\rm UV}(k^2,p^2)]^2\theta(k^2-p^2).
\label{G1approx2}
\end{eqnarray}
As was discussed in Sec.~\ref{sec_ren_hyp},
the ultraviolet channel function $F_{\rm UV}(k^2,p^2)$ is proportional 
to $(\Lambda/p)^{\eta/2}$ and $p^2\Delta_{S}(p)$ is proportional to 
$(\Lambda/p)^{-\eta}$. Therefore, the cutoff dependence cancels 
in Eq.~(\ref{G1approx2}) as was expected and the angular integral 
Eq.~(\ref{G1def}) can indeed be written in terms of a function which 
depends only on the ratio of $p^2/k^2$.

The scaling form ($p^2\ll \Lambda^2$)
for the scalar propagator $\Delta_{S}(p)$ at $g_0=g_c$ ($m_\sigma=0$) is 
given by Eq.~(\ref{scalhypscalprop}) and Eq.~(\ref{Bw}).
The scaling form for the channel function $F_{\rm UV}(k^2,p^2)$ 
is given by Eqs.~(\ref{scalfuv1}) 
and (\ref{scalfuv2}),
with $p^2\leq k^2\ll \Lambda^2$.
Inserting Eqs.~(\ref{scalhypscalprop}) (with $m_\sigma=0$), 
(\ref{scalfuv1}), and (\ref{scalfuv2}) in Eq.~(\ref{G1approx2}), 
we obtain for $G_1$ 
\begin{eqnarray}
G_1(u)&=&\frac{\Gamma(2-\omega)
\Gamma(2+\omega)}{8\omega \gamma(\omega)\gamma(-\omega)}
\nonumber\\&\times&
u\left[\gamma(\omega)I_{-\omega}\left(\sqrt{2\lambda_0 u}\right)
-\gamma(-\omega)I_{\omega}\left(\sqrt{2\lambda_0 u}\right)\right]^2
\theta(1-u),\label{G1esexp}
\end{eqnarray}
where $u=p^2/k^2$.
Thus Eq.~(\ref{contlimphi1}) is
\begin{eqnarray}
\phi_1(\alpha_0,g_c)&\approx&-N\zeta_1(\alpha_0),\label{phi1zeta1}\\
\zeta_1(\alpha_0)&=&\int\limits_0^1{d}u\, G_1(u) \geq 0.\label{zeta1}
\end{eqnarray}
The function $G_1$ is positive, hence $\phi_1$
is negative.
The integral over the function $G_1$ can be done explicitly by making 
use of the integral identity 2.15.19.1 in Volume 2 of 
Prudnikov {\em et al.} \cite{prbrma86}.
The result is
\begin{eqnarray}
\zeta_1(\alpha_0)&=&\frac{1}{\omega}\frac{\lambda_0}{2}
\Biggr\{\frac{1}{(2+\omega)}
\frac{\Gamma(1-\omega) \gamma(-\omega)}{\Gamma(1+\omega)\gamma(\omega)}
\left(\frac{\lambda_0}{2}\right)^\omega\nonumber\\
&\times&{_2}F_3\left(2+\omega,1/2+\omega;
3+\omega,1+2\omega,1+\omega;2\lambda_0\right)\nonumber\\
&-&{_2}F_3\left(2,1/2;
3,1+\omega,1-\omega;2\lambda_0\right)\nonumber\\
&+&\frac{1}{(2-\omega)}
\frac{\Gamma(1+\omega) \gamma(\omega)}{\Gamma(1-\omega)\gamma(-\omega)}
\left(\frac{\lambda_0}{2}\right)^{-\omega}\nonumber\\
&\times&{_2}F_3\left(2-\omega,1/2-\omega;
3-\omega,1-2\omega,1-\omega;2\lambda_0\right)\Biggr\}.\label{zeta1expl}
\end{eqnarray}

The above analysis of the function $\phi_1$ is repeated for the function 
$\phi_3$. 
The second derivative, $K_\alpha^{(2)\alpha}(p,k)$ (see Eq.~(\ref{k2derdef})),
of the BS kernel given by Eq.~(\ref{K2approx}) is 
\begin{eqnarray}
K_\alpha^{(2)\alpha}(p,k)
\propto \lim_{q\rightarrow 0}
\frac{\partial^2}{\partial q^\alpha \partial q_\alpha}F_1(k+q,p+q). 
\end{eqnarray}
The SDE for $F_1$ (in quenched-ladder approximation, 
see Fig.~\ref{fig_scalarvertexladder}) is 
given by 
\begin{eqnarray}
F_1(k+q,p+q)&=&1-i
\lambda_0\int_\Lambda\frac{{d}^4r}{\pi^2}\,
\frac{(r^2+(k-p)\cdot r)}{r^2(r+k-p)^2(r-p-q)^2} F_1(r+k-p,r),
\end{eqnarray}
where we recall that we neglect the vertex function $F_2$.
Thus
\begin{eqnarray}
\lim_{q\rightarrow 0}
\frac{\partial^2}{\partial q^\alpha \partial q_\alpha}F_1(k+q,p+q)
&=&
-i\lambda_0\int_\Lambda\frac{{d}^4r}{\pi^2}\,
\frac{(r^2+(k-p)\cdot r)}{r^2(r+k-p)^2}F_1(r+k-p,r)\nonumber\\
&\times&\lim_{q\rightarrow 0}
\frac{\partial^2}{\partial q^\alpha \partial q_\alpha}
\frac{1}{(r-p-q)^2}. \label{laplsde}
\end{eqnarray}
By making use of the identity
\begin{eqnarray}
\frac{\partial}{\partial q^\alpha}
\frac{\partial}{\partial q_\alpha}\frac{1}{q^2}
=-4\pi^2i \delta^4(q),
\end{eqnarray}
we obtain
\begin{eqnarray}
{K_\alpha^{(2)\alpha}}_{^{cd,ab}_{kl,ij}}(p,k)&=&
\frac{\delta_{ij}\delta_{kl}}{e_0^2}
 \left(-4\lambda_0\frac{p\cdot k}{p^2 k^2}\right)
\left[F_1(p,k)\right]^2 \Delta_{S}(p-k)
\left[{\bf 1}_{ad}{\bf 1}_{cb}+{i{\gamma_5}}_{ad}{i{\gamma_5}}_{cb}\right],
\end{eqnarray}
in Minkowsky formulation.
Inserting the above expression in Eq.~(\ref{phi3})
the equation for $\phi_3$ takes the form (in Euclidean formulation)
\begin{eqnarray}
\phi_3(\alpha_0,g_c)=-\lim_{\Lambda\rightarrow \infty}\frac{\alpha_0}{\pi}
\frac{N}{8\pi^2}
\int\limits_0^{\Lambda^2}
{d}p^2\,\int \frac{{d}\Omega_p}{2\pi^2}\,
\frac{p^2}{k^2}\frac{(p\cdot k+k^2)^3}{(p+k)^6}
[F_1(k+p,k)]^2\Delta_{S}(p).
\end{eqnarray}
Then 
\begin{eqnarray}
\phi_3(\alpha_0,g_c)
&=& N\int\limits_0^{\infty}{d}u\, G_3(u),
\label{contlimphi3}\\
G_3(p^2/k^2)&\equiv&
\lim_{\Lambda\rightarrow \infty}
-\frac{\alpha_0}{\pi}\frac{1}{8\pi^2} 
\int \frac{{d}\Omega_p}{2\pi^2}\,
\frac{p^2(k\cdot p+k^2)^3}{(k+p)^6}[F_1(k+p,k)]^2\Delta_{S}(p).\label{G3def}
\end{eqnarray}
We use the following Chebyshev expansion:
\begin{eqnarray}
\frac{(k\cdot p+k^2)^3}{(k+p)^6}
=\sum_{n=0}^\infty d_n(k^2,p^2) U_n(\cos\alpha),\quad 
\cos\alpha=\frac{k\cdot p}{kp},
\end{eqnarray}
where
\begin{eqnarray}
d_n(k^2,p^2)&=&\frac{2}{\pi}\int\limits_0^\pi{d}\alpha\,
\sin^2\alpha\,U_n(\cos\alpha)\frac{(p\cdot k+k^2)^3}{(p+k)^6},\\
d_0(k^2,p^2)&=&\left(1-\frac{3p^2}{4k^2}\right)\theta(k^2-p^2),\label{d0pr}\\
d_n(k^2,p^2)&=& \frac{(-1)^n}{8}
\left\{n+1+\left[6+\sum_{l=0}^{n-1}(4+l)\right] 
\left(1-\frac{p^2}{k^2}\right)\right\}
\theta(k^2-p^2)\left(\frac{p}{k}\right)^n \nonumber\\
&-&\frac{(-1)^n}{8}
\left\{n+1-\left[\sum_{l=0}^{n-1}(2-l)\right]\left(1-\frac{k^2}{p^2}\right)
\right\}\theta(p^2-k^2)\left(\frac{k}{p}\right)^n,\qquad n\geq 1.
\end{eqnarray}
The function $G_3$ can be expressed as
\begin{eqnarray}
G_3(p^2/k^2)=\lim_{\Lambda\rightarrow \infty}
-\frac{\alpha_0}{\pi}\frac{p^2\Delta_{S}(p)}{8\pi^2} 
\sum_{l,m,n=0}^\infty C_{lmn} \,d_l(k^2,p^2) f_m(k^2,p^2) f_n(k^2,p^2).
\label{G3exp}
\end{eqnarray}
We also approximate $G_3$ by keeping only the
lowest order term in the Chebyshev expansion,
\begin{eqnarray}
G_3(p^2/k^2)\approx\lim_{\Lambda\rightarrow \infty}
-\frac{\alpha_0}{\pi}\frac{p^2\Delta_{S}(p)}{8\pi^2} 
d_0(k^2,p^2) \left[f_0(k^2,p^2)\right]^2. \label{G3approx}
\end{eqnarray}
Then, again the asymptotics, $k^2\gg p^2$,  respectively, $p^2\gg k^2$,
of $G_3$ are well approximated by the lowest order Chebyshev term. 
Moreover, for momenta $k^2=p^2$ 
the approximation Eq.~(\ref{G3approx}) is exact, since
$d_n(k^2,k^2)=0$ for all $n\geq 1$.
Therefore, the approximation Eq.~(\ref{G3approx}) is even better
than the analogous approximation, Eq.~(\ref{G1approx}), to $G_1$. 
Furthermore, from Eq.~(\ref{d0pr}) we see that the only nonzero contributions 
to $G_3$ of Eq.~(\ref{G3approx}) 
are given by momenta $k^2\geq p^2$. Thus, using Eqs.~(\ref{G3approx}),
(\ref{channelapprox}), and (\ref{d0pr}), we find
\begin{eqnarray}
\!\!\!\!
G_3(p^2/k^2)&\approx& 
\lim_{\Lambda\rightarrow \infty}
-\frac{\alpha_0}{\pi}\left(1-\frac{3p^2}{4k^2}\right)
\frac{p^2\Delta_{S}(p)}{8\pi^2} [F_{\rm UV}(k^2,p^2)]^2
\theta(k^2-p^2).\label{G3approx2}
\end{eqnarray}
Substituting Eqs.~(\ref{scalhypscalprop}), (\ref{scalfuv1}), and 
(\ref{scalfuv2}) in Eq.~(\ref{G3approx2}),
we obtain for $G_3$
\begin{eqnarray}
G_3(u)&=&\frac{\alpha_0}{\pi}\left(1-\frac{3u}{4}\right)\frac{\Gamma(2-\omega)
\Gamma(2+\omega)}{8\omega \gamma(\omega)\gamma(-\omega)}
\nonumber\\&\times&u\left[\gamma(\omega)
I_{-\omega}\left(\sqrt{2\lambda_0 u}\right)
-\gamma(-\omega)I_{\omega}\left(\sqrt{2\lambda_0 u}\right)\right]^2
\theta(1-u),\label{G3esexp}
\end{eqnarray}
where $u=p^2/k^2$.
Thus Eq.~(\ref{contlimphi3}) is
\begin{eqnarray}
\phi_3(\alpha_0,g_c)&\approx& N\zeta_3(\alpha_0),\label{phi3zeta3}\\
\zeta_3(\alpha_0)&=&\int\limits_0^1{d}u\, G_3(u) \geq 0.\label{zeta3}
\end{eqnarray}
The function $\phi_3$ is positive, and can be computed in the same way as 
$\phi_1$. The result is
\begin{eqnarray}
\zeta_3(\alpha_0)=\frac{\alpha_0}{\pi}\left[\zeta_1(\alpha_0)
-\tau(\alpha_0)\right],\label{zeta3expl}
\end{eqnarray}
where
\begin{eqnarray}
\tau(\alpha_0)&=&\frac{3}{4\omega}\frac{\lambda_0}{2}
\Biggr\{\frac{1}{(3+\omega)}
\frac{\Gamma(1-\omega) \gamma(-\omega)}{\Gamma(1+\omega)\gamma(\omega)}
\left(\frac{\lambda_0}{2}\right)^\omega\nonumber\\
&\times&{_2}F_3\left(3+\omega,1/2+\omega;
4+\omega,1+2\omega,1+\omega;2\lambda_0\right)\nonumber\\
&-&\frac{2}{3}{_2}F_3\left(3,1/2;
4,1+\omega,1-\omega;2\lambda_0\right)\nonumber\\
&+&\frac{1}{(3-\omega)}
\frac{\Gamma(1+\omega) \gamma(\omega)}{\Gamma(1-\omega)\gamma(-\omega)}
\left(\frac{\lambda_0}{2}\right)^{-\omega}\nonumber\\
&\times&{_2}F_3\left(3-\omega,1/2-\omega;
4-\omega,1-2\omega,1-\omega;2\lambda_0\right)\Biggr\}.\label{taueq}
\end{eqnarray}

In the computation of the functions $\phi_1$, $\phi_2$, and 
$\phi_3$ the ladder (planar) photon exchanges have been neglected.
After reinstating the ladder photon exchange term 
of Eq.~(\ref{K2approx}), 
we obtain, together with Eqs.~(\ref{phi1zeta1}) and (\ref{phi3zeta3}), 
that
\begin{eqnarray}
\phi_1(\alpha_0,g_c(\alpha_0))=\frac{\alpha_0}{2\pi}-N\zeta_1(\alpha_0),
\quad \phi_2(\alpha_0,g_c(\alpha_0))=0,
\quad\phi_3(\alpha_0,g_c(\alpha_0))
=N\zeta_3(\alpha_0).\label{phi1phi3phot}
\end{eqnarray}
The ladder photon exchange only contributes to $\phi_1$, see again
\cite{jowiba67}.
After substitution of Eq.~(\ref{phi1phi3phot}) in Eq.~(\ref{betasc2a}),
the $\beta$ function reads
\begin{eqnarray}
\beta_\alpha(\alpha_0,g_c)=\frac{N\alpha_0^2}{\pi}
\left[\frac{2}{3}+\frac{\alpha_0/2\pi-N\zeta_1(\alpha_0)}{1-\alpha_0/2\pi
+N\zeta_1(\alpha_0)}+N\zeta_3(\alpha_0)\right],\label{betasc2}
\end{eqnarray}
where explicit expressions for $\zeta_1$ and $\zeta_3$ 
are given by Eq.~(\ref{zeta1expl}) and Eqs.~(\ref{zeta3expl}) and (\ref{taueq}).
\section{UV stable fixed points}\label{critfixpoints}
\begin{figure}[t!]
\epsfxsize=10cm
\epsffile[120 320 370 560]{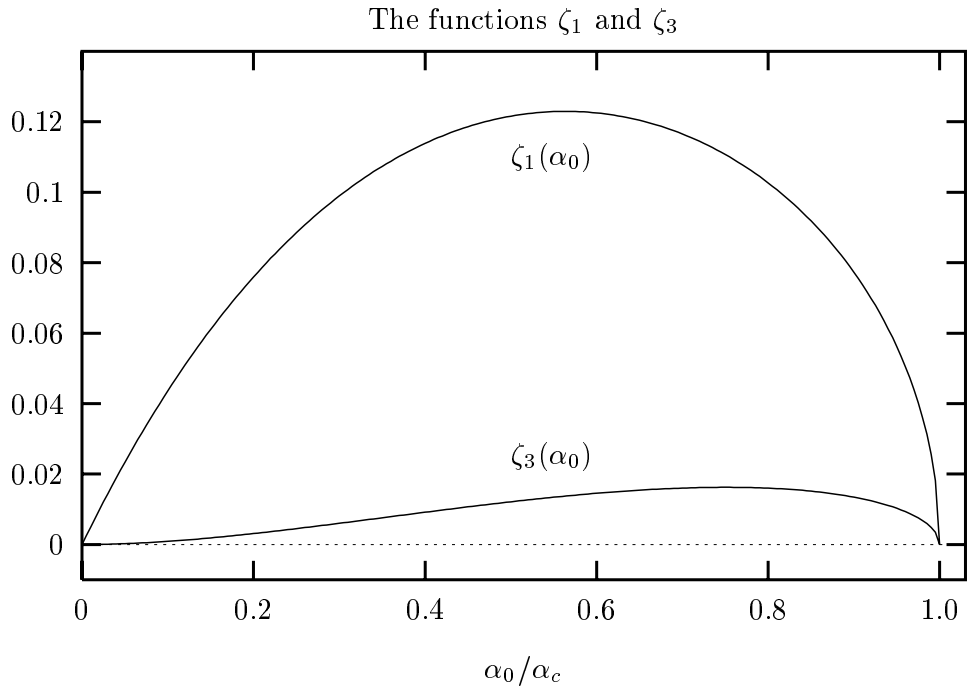}
\caption{The functions $\zeta_1$ and $\zeta_3$ plotted versus
$\alpha_0/\alpha_c$.}
\label{fig_zeta13}
\end{figure}

Let us start analyzing Eq.~(\ref{betasc2}) by first considering the
properties of the functions $\zeta_1(\alpha_0)$ and $\zeta_3(\alpha_0)$. 
For $\alpha_0$ small, the expansion of the functions $\zeta_1$ and $\zeta_3$ 
can be computed from Eqs.~(\ref{zeta1expl}) and (\ref{zeta3expl}). The 
result is 
\begin{eqnarray}
\zeta_1(\alpha_0)\approx\frac{3\alpha_0}{2\pi}+{\cal O}(\alpha_0^2),\qquad
\zeta_3(\alpha_0)\approx\frac{15}{16}
\frac{\alpha_0^2}{\pi^2}+{\cal O}(\alpha_0^3),\label{z1z3exp}
\end{eqnarray}
showing that $\zeta_3$ vanishes faster that $\zeta_1$ for 
$\alpha_0\rightarrow 0$. 
The functions $\zeta_1(\alpha_0)$ and $\zeta_3(\alpha_0)$
have been plotted versus $\alpha_0/\alpha_c$ in Fig.~\ref{fig_zeta13}. 
First, it is clear that $\zeta_1$ and $\zeta_3$, are positive, and 
have a maximum at some intermediate value of $0<\alpha_0<\alpha_c=\pi/3$. 
For instance, $\zeta_1$ has a maximum $\zeta_1\approx 0.123$ 
at $\alpha_0/\alpha_c\approx 0.58$ ($\omega\approx 0.65$).
Second, the functions $\zeta_1$ and $\zeta_3$ vanish at the pure NJL 
point $\alpha_0=0$ in accordance with Eq.~(\ref{z1z3exp}), 
and at the CPT point $\alpha_0=\alpha_c$.
At $\alpha_0=0$, we can consider this is as a reflection of the fact
that hyperscaling breaks down due to logarithmic corrections; 
the ``effective'' Yukawa coupling is trivial, therefore vanishes.
At $\alpha_0=\alpha_c$, where the critical exponents 
become singular, the vanishing of $\zeta_1$ and $\zeta_3$
is related to the dynamics of the conformal phase transition (CPT), 
which has been thoroughly discussed in Ref.~\cite{miya97}. 
There are no light $\sigma$ and $\pi$ 
exchanges in the symmetric phase \cite{gure98}
which consequently implies the absence of
effective Yukawa interactions.\footnote{Moreover at the CPT point four-fermion
interaction are marginal instead of relevant, and start to mix
with the gauge interaction, hence the analysis becomes considerably
more complicated.}

Let us compare the $\beta$ function (\ref{betasc2})
with the $\beta$ function (\ref{betscal}) of the gauge--Higgs--Yukawa 
model (\ref{GHY}) in the $1/N$ expansion. 
Then, the entire set of planar $\sigma$ and $\pi$ exchanges
is generated by the kernel
\begin{eqnarray}
K^{(2)}_{^{cd,ab}_{kl,ij}}(p,p+q,k+q)
\approx\delta_{ij}\delta_{kl}\frac{g_Y^2}{e_0^2}\Delta_{S}(k-p) 
\left[{\bf 1}_{ad}{\bf 1}_{cb}+{i{\gamma_5}}_{ad}{i{\gamma_5}}_{cb}\right],
\label{K2approxGHY}
\end{eqnarray}
where $\Delta_{S}(p)=1/p^2$. 
With such a kernel, $\phi_2$ and $\phi_3$ are zero, because the
right-hand side of Eq.~(\ref{K2approxGHY}) does not depend on the 
momentum $q$.
The expression for $\phi_1$, in this case, can be computed straightforwardly
($\lambda_Y=g_Y^2/4\pi$);
\begin{eqnarray}
\phi_1&=&\lim_{\Lambda\rightarrow \infty}
2 N g_Y^2 i\int_\Lambda\frac{{d}^4p}{(2\pi)^4}
\frac{p\cdot k}{p^4} \frac{1}{(p-k)^2}=-\frac{N \lambda_Y}{2\pi}.
\end{eqnarray}
Again we introduce the ladder photon exchanges by the
replacement
\begin{eqnarray}
\phi_1(\lambda_Y)\quad \longrightarrow \quad
\phi_1(\alpha_0,\lambda_Y)=\frac{\alpha_0}{2\pi}-\frac{N \lambda_Y}{2\pi}.
\end{eqnarray}
Hence, in this case, the $\beta$ function is
\begin{eqnarray}
\beta_\alpha(\alpha_0,\lambda_Y)=\frac{N\alpha_0^2}{\pi}
\left[\frac{2}{3}
+\frac{\alpha_0/2\pi-N \lambda_Y/2\pi}{1-\alpha_0/2\pi
+N \lambda_Y/2\pi}\right].\label{betajwbGHY}
\end{eqnarray}
Comparing the $\beta$ functions (\ref{betasc2}) and (\ref{betajwbGHY})
leads to the suggestion that $\zeta_1(\alpha_0)$ is analogous to the 
Yukawa coupling $\lambda_Y$ in a gauge-Higgs-Yukawa model, 
$\zeta_1(\alpha_0)\sim \lambda_Y/2\pi$.

This is a crucial point.
The general consensus is that for a gauge-Higgs-Yukawa model 
the Yukawa interaction $\lambda_Y$ is trivial, thus
$\lambda_Y\rightarrow 0$ in Eq.~(\ref{betajwbGHY}). 
However, the situation is essentially different for $\zeta_1$ in the GNJL 
model.
There the ``effective'' coupling $\zeta_1$ 
is formed by the exchange of $\sigma$ and $\pi$ bosons, with
the Yukawa vertices, and (pseudo)scalar propagators fully
dressed ({\em i.e.}, the skeleton expansion). 
The cancellation of the $Z$ factors, see Sec.~\ref{sec_ren_hyp},
which is related to the fact that the hyperscaling equations are satisfied,
gives rise to a finite nonzero $\zeta_1(\alpha_0)$ at the
critical curve ($g_0=g_c$) for $0<\alpha_0<\alpha_c$.
The other nonzero function $\zeta_3$ results from taking into account fully 
dressed Yukawa vertices.

Let us now the discuss the possible existence of UV stable fixed points.
A necessary but not a sufficient condition for the realization of
an UV stable fixed point is that $N\zeta_1$ has to be larger than both 
$N\zeta_3$ and $\alpha_0/2\pi$, and $N\zeta_1\sim {\cal O}(1)$.
For large $N$, the contribution of the planar photon exchanges 
(represented by the $\alpha_0/2\pi$ terms) is negligible
with respect to $N\zeta_1$ and $N\zeta_3$.
Moreover Fig.~\ref{fig_zeta13} shows, for $\alpha_0$ small, that 
$\zeta_1$ is considerably larger than $\zeta_3$.
This means that only for flavors $N$ larger than some critical value 
$N_c$ UV stable fixed points can be obtained.

\begin{figure}[b!]
\epsfxsize=10cm
\epsffile[120 320 370 560]{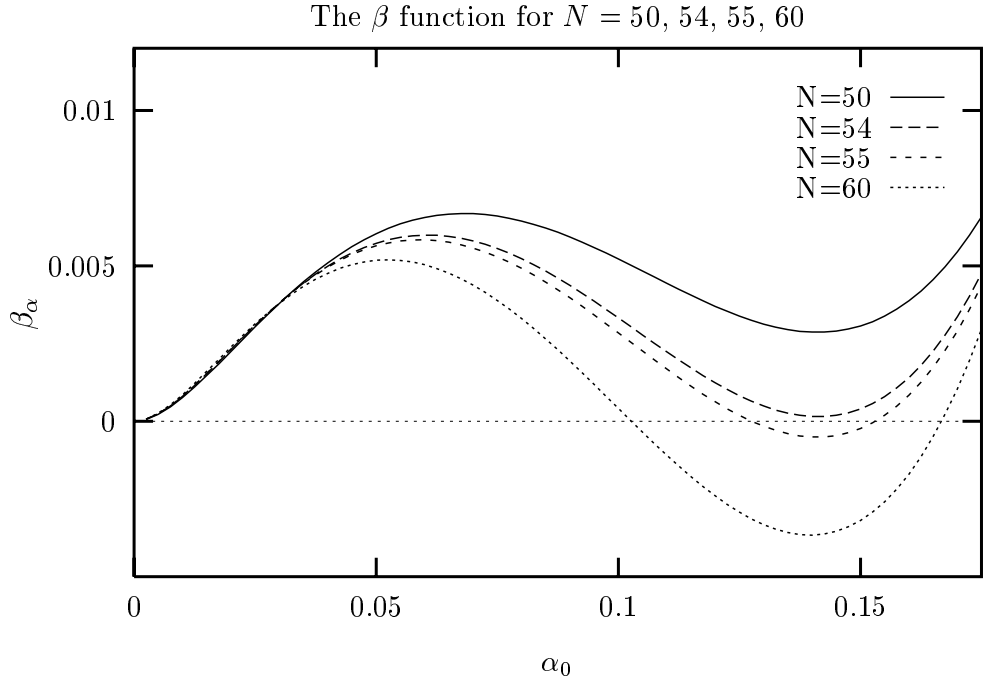}
\caption{Plot of the function $\beta_\alpha$ versus the gauge coupling 
$\alpha_0$ for various values of the fermion flavor number $N$.}
\label{fig_beta5060}
\end{figure}

By substituting the expressions (\ref{zeta1expl}) and (\ref{zeta3expl})
for $\zeta_1$ and $\zeta_3$ in Eq.~(\ref{betasc2}), 
we can straightforwardly analyze the $\beta$ function graphically.
In Fig.~\ref{fig_beta5060} the $\beta$ function 
is plotted for various values of $N$.
Figure~\ref{fig_beta5060} shows that for values of 
$N>N_c$, with $55>N_c>54$, UV stable fixed points 
exist, the largest being $\alpha_\star\approx 0.13$;
\begin{eqnarray}
N=55:&\quad&\beta_\alpha\left(0.13,g_c(0.13)\right)\approx 0,\quad 
\eta_\alpha=-\beta_\alpha^\prime\left(0.13,g_c(0.13)\right)\approx
0.07,\\
N=60:&\quad&\beta_\alpha\left(0.1,g_c(0.1)\right)\approx 0,\quad
\eta_\alpha=-\beta_\alpha^\prime\left(0.1,g_c(0.1)\right)\approx
0.15.
\end{eqnarray}
In accordance with Eqs.~(\ref{UVfixpointdef}) and (\ref{UVfixdef2}), 
the fixed points are first-order zeros of $\beta$.
The general pattern is clear;  
the larger $N$, with $N>N_c$, the smaller will be the UV stable fixed point, 
but the larger will be the critical exponent $\eta_\alpha$.

The above pattern also suggests that when $N\rightarrow \infty$ 
the UV stable fixed point $\alpha_\star\rightarrow 0$ 
and we would obtain an asymptotically free theory.
This is not the case. 
It was shown in Refs.~\cite{aptewij91,gure98} as $\alpha_0$ goes to zero that
a logarithmic correction appears in the expression for the scalar propagator
$\Delta_{S}$.
In fact, the scaling form for
$\Delta_{S}$ (with $q/\Lambda\ll 1$) is only 
valid for values of $\omega$ 
so that $(q^2/\Lambda^2)^\omega\gg q^2/\Lambda^2$, see Ref.~\cite{gure98}.
The logarithmic correction gives rise to the breakdown of hyperscaling 
relations and is synonymous to triviality of the four-fermion interactions
(the NJL model). 
Since our results rely heavily on the existence of scaling forms
such as Eqs.~(\ref{scalhypscalprop}) and (\ref{scalfuv1}),
we can only trust our results for values of $\alpha_\star$ not too small. 

Since we have made use of results obtained in the quenched approximation, 
we mention that the plots of the $\beta$ function are (at the most) 
reliable at or in the vicinity of the UV stable fixed points 
at which the quenched approach is self consistent, 
see Eq.~(\ref{quenchhyp2}).

\begin{figure}[b!]
\epsfxsize=10cm
\epsffile[120 320 370 560]{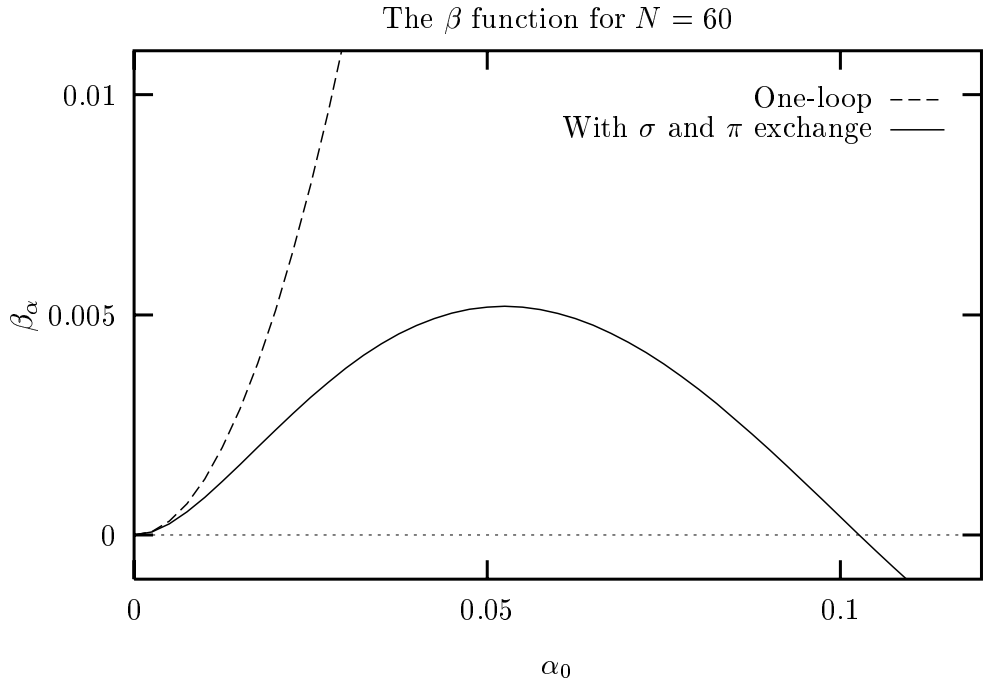}
\caption{The one-loop $\beta$ function for $N=60$
compared with the $\beta$ function including four-fermion interactions.}
\label{fig_beta60}
\end{figure}

In Fig.~\ref{fig_beta60}, the case of $N=60$ fermion flavors is 
compared with the one-loop $\beta$ function of QED.
For very small values of $\alpha_0< 1/100$ indeed the one-loop QED result
coincides with that of the GNJL model, however for larger values of $\alpha_0$
the $\beta$ function (\ref{betasc2}) deviates from the one-loop 
expression, and eventually an UV stable fixed point is realized 
at $\alpha_\star\approx 0.1$.

The analysis shows that a rather large number of flavors, 
\begin{eqnarray}
N>N_c\approx 54, 
\end{eqnarray}
is
required to obtain UV stable fixed points.
From the point of view of the $1/N$ expansion this seems a consistent
result, since other than planar contributions are suppressed 
by at least factors of (say) $1/N_c$. 
However, from the phenomenological point of view,
the result is unsatisfactory, since it implies that the unquenched Abelian
GNJL model (exhibiting UV stable fixed points) is only practically 
applicable for 
models which have at least $N_c$ fermion flavors (fractions rounded up). 
Therefore, it is appropriate to discuss how $N_c$ depends on the 
approximation.

First, we stress that
the second term on the right-hand side
in Eq.~(\ref{betasc2}) containing the $\zeta_1$ function
causes the suppression of charge screening and
is responsible for the possible realization of an UV stable fixed point.
The denominator in the second term 
is a direct consequence of the resummation of the infinite ladder $\sigma$
and $\pi$ exchanges, and it is mainly due to this
denominator $1+N\zeta_1$ that the critical number of fermion flavors
$N_c$ is large.

Second, the existence of an UV stable fixed point for a specific 
number of fermion flavors $N$ depends on the interplay between 
the functions $\zeta_1$ and $\zeta_3$, which are given in terms of integrals
of the functions $G_1$ and $G_3$. 
Let us recall that the lowest order Chebyshev expansion
for $G_1$, Eq.~(\ref{G1approx}) is a lower bound on $G_1$ 
of Eq.~(\ref{G1exp}), since all terms of the Chebyshev 
expansion are positive at $k^2=p^2$, the same cannot be said about the
approximation (\ref{G3approx}) for $G_3$.
Thus keeping more terms in the Chebyshev expansion leads to an 
increase of $\zeta_1$, whereas the effect on $\zeta_3$ is less clear,
because of the alternating Chebyshev series for $\zeta_3$. 
Therefore, an improvement of the computation of $\zeta_1$ 
will probably lead to a decrease of the critical flavor number $N_c$.

Moreover, in the computation of the functions $\zeta_1$ and $\zeta_3$ 
we have used Yukawa vertices (${\Gamma_{S}}$) and $\sigma$ and 
$\pi$ propagators ($\Delta_{S}$) which were obtained in the quenched-ladder 
approximation.
An interesting question is whether the improvement
of the ladder approximation for the gauge interaction
({\em e.g.}, by including crossed photon exchanges)
leads to a increase of $\zeta_1$, and thus a decrease of $N_c$.

Finally, we recall that we have neglected 
the effect of the Yukawa vertex function $F_2$ (Eq.~(\ref{vertfiesdef})),
but clearly the inclusion of $F_2$ in the analysis 
could change the results quantitatively. 
Whether such an improvement will tend to increase or decrease $N_c$ remains 
unclear at this stage.
\section{Conclusion}\label{sec_concl}
There are strong indications that four-fermion interactions become 
relevant near the chiral phase transition in GNJL models in four dimensions,
due to the appearance of a large anomalous dimension. 
The main objective of this paper was to study the effect of such relevant
four-fermion interactions on the vacuum polarization of the gauge coupling
and to reinvestigate the problem of triviality for a particular 
Abelian GNJL model with $N$ fermion flavors.
To obtain new results, the four-fermion interactions had to
be taken into account 
beyond the commonly used Hartree--Fock or mean-field approach.

The crucial feature of the GNJL model, 
within the quenched mean-field approximation, is that 
a nontrivial Yukawa interaction ({\em i.e.}, an interaction between 
composite (pseudo)scalars and fermions) exists for $0<\alpha_0<\alpha_c$.
The existence of such a nontrivial Yukawa interaction
requires the cancellation of 
renormalization constants of the $\sigma$ and $\pi$ fields
in fermion-antifermion scattering amplitudes such as the BS 
kernel $K^{(2)}$. This is analogous to the requirement of hyperscaling 
(see Sec.~\ref{sec_ren_hyp}).
If the hyperscaling equations are satisfied,
then only two of the critical exponents are independent, {\em e.g.},
$\eta$ and $\gamma$, see Eqs.~(\ref{delbetgamnu}) and (\ref{etaexpr}).
The existence of hyperscaling relations between the critical exponents
is intimately connected with the existence of Ward--Takahashi identities
(and thus the Goldstone mechanism) arising from the continuous symmetries 
of the model.

The skeleton expansion for the BS kernel $K^{(2)}$ provides a natural 
framework to take into account the anomalous dimensions of Yukawa 
vertices and $\sigma$ and $\pi$ propagators.
Within the skeleton expansion, $\sigma$ and $\pi$ exchanges are described
in terms of fully dressed Yukawa vertices and $\sigma$ 
and $\pi$ propagators.
The actual computation of the anomalous dimension, and the resolution of 
the scaling form requires a solution of the SDE's for 
Yukawa vertices, and $\sigma$ and $\pi$ bosons. 

In previous work such fully dressed Yukawa vertices and $\sigma$ and $\pi$ 
propagators have been analyzed in the quenched-ladder mean-field
approximation, see Ref.~\cite{gure98} and references therein.
To make use of these results consistently, we used the following
approximations.
First, we assumed that the bare coupling parameters are fine-tuned
close to the critical point, {\em i.e.}, close to an UV stable fixed point, 
at which $\beta_g\approx 0$ and $\beta_\alpha\approx 0$.
In that case, the quenched or canonical approximation for the photon 
propagator is self-consistent.
Second, the gauge-interaction is considered in the ladder form, with bare 
vertices.
Third, we used the $1/N$ expansion (with $N$ the number of fermion flavors)
which states that planar $\sigma$ and $\pi$ exchanges describe the
leading contribution to Green functions for large $N$.
Then, due to the specific form of the chiral symmetry
with both scalars and pseudoscalar in the adjoint representation of the
$U_L(N)\times U_R(N)$ symmetry, we argued that in so-called zero-spin 
channels (such as Yukawa vertices and $\sigma$ and $\pi$ propagators) 
the planar $\sigma$ and $\pi$ exchanges cancel each other 
for momenta larger than the mass of the $\sigma$ boson 
(in fact in the symmetric phase this cancellation is exact).
Moreover, an important property of the planar (ladder) approximations
is that they respect the vector and chiral Ward--Takahashi 
identities.

The method of Ref.~\cite{jowiba67}, provides a nonperturbative 
framework independent of the fermion wave function ${\cal Z}$, and
allowed us to compute the contributions of the infinite set of 
planar $\sigma$ and $\pi$ exchanges to the vacuum polarization. 
The result of the computations is that the GNJL model exhibits an UV stable 
fixed point, $\beta_\alpha(\alpha_\star,g_c(\alpha_\star))=0$, 
for any value of $N$ that exceeds some critical 
value $N_c$ ($N>N_c$).
This critical number of flavors turned out to be $N_c\approx 54$.
The larger the number of fermion flavors, 
the smaller the UV stable fixed point $\alpha_\star$ will be, 
provided $N>N_c$ and $\alpha_\star$ not too small.
Since our results are derived on the basis of the existence
of hyperscaling laws, we cannot extrapolate our results
into the region where hyperscaling breaks down due to 
logarithmic violations, {\em i.e.}, when $\alpha_\star\rightarrow 0$ 
($N\rightarrow \infty$).  

From a phenomenological point of view, 
the large value for $N_c$ puts questions to the applicability of the GNJL.
However, we have given a few arguments in the previous section suggesting 
that $N_c$ could be rather sensitive to approximations,
and that an improvement of the approximations and calculations 
will probably lead to a smaller value for $N_c$.

The realization of an UV stable fixed point 
is motivated by the observation that contributions of
planar $\sigma$ and $\pi$ exchanges to the vacuum polarization, 
in an Abelian gauge--Higgs--Yukawa model,
have identical sign, and tend to reduce screening. 
In analogy, four-fermion interactions describe attractive forces
between virtual fermion-antifermion pairs in the vacuum polarization.

The conventional leading term in the vacuum polarization is 
the one-loop correction describing the creation of  
fermion-antifermion pairs. These virtual pairs can be considered as 
dipoles causing the screening; the vacuum is a medium of the insulator type.
Such a screening is proportional to the coupling 
$\alpha_0$ and proportional to the number of fermion flavors $N$.
However, if a particular fraction of the total 
amount of fermion-antifermion pairs created
are correlated by attractive four-fermion interactions, represented by
$\sigma$ and $\pi$ exchanges, then clearly these
composite neutral states are not capable of screening.
The negative term $N\zeta_1$ in the $\beta$ function (\ref{betasc2}) 
represents the contributions and the attractive nature of 
four-fermion interactions in the vacuum polarization.

Within the quenched-ladder mean-field approximation, 
the critical curve and critical exponents are independent of the number 
of fermion flavors.
Within our approximation scheme, the mechanism of charge screening 
clearly is flavor dependent, since the
total number of virtual fermion-antifermion pairs is proportional to $N$ 
and the total number of composite scalars and pseudoscalars grows as $2N^2$.
The larger the number of flavors, the stronger the effect of four-fermion
interactions. 
The fixed point appears when the virtual pairs completely loose
their ability to screen. 

The existence of an UV stable fixed point implies a nontrivial continuum 
limit of the Abelian GNJL model.
The analysis presented here suggest that in the full unquenched GNJL model 
the critical line is replaced by 
an UV stable fixed point (on the critical line) whose exact positions depends 
on the number of fermion flavors.
If the number of fermion flavors is below some specific value, the critical
four-fermion dynamics are not sufficient to yield an UV stable fixed point. 
In that case the unquenched GNJL model only has a trivial (IR) fixed point
and the chiral phase transition is of the mean-field type.
\acknowledgements{The author wishes to thank Valery Gusynin 
for the enjoyable and fruitful
collaboration, the numerous stimulating discussions, and important 
suggestions.
It is a pleasure to thank Marinus Winnink for useful comments and
encouragement.}
\appendix
\section{Two-loop vacuum polarization}\label{appvacpol}
In this appendix we compute 
two-loop vacuum polarization corrections including $\sigma$ and $\pi$ 
exchanges, see Fig.~\ref{fig_twoloop}. 
We derive the two-loop contribution by making use of the 
one-loop computation of the photon-fermion vertex \cite{bach80,kirepe95}.

The SDE for vacuum polarization tensor reads ($N=1$)
\begin{eqnarray}
\Pi^{\mu\nu}(q^2)=
ie_0^2\int_\Lambda\frac{d^4k}{(2\pi)^4}\,
{\rm Tr}\left[\gamma^\mu S(k+q) \Gamma^\nu(k+q,k) S(k)\right].
\end{eqnarray}
Assuming that the WTI's are respected, 
the vacuum polarization tensor is transverse: 
$\Pi^{\mu\nu}(q)=\left(-g^{\mu\nu}q^2+q^\mu q^\nu\right) \Pi(q^2)$,
so that 
\begin{eqnarray}
\Pi(q^2)=
-\frac{ie_0^2}{3q^2}\int_\Lambda\frac{d^4k}{(2\pi)^4}\,
{\rm Tr}\left[\gamma_\mu S(k+q) \Gamma^\mu(k+q,k) S(k)\right].
\end{eqnarray}
Let us write and denote the one-loop vertex and self-energy corrections
with a subscript $(1)$ as follows:
\begin{eqnarray}
\Gamma^\mu(k,p)=\gamma^\mu+\Lambda^{\mu}_{(1)}(k,p),\qquad 
S(p)=\frac{\hat p}{p^2}\left[1+{\cal Z}_{(1)}(p^2)\right]. \label{1loopcors}
\end{eqnarray}
Besides a photon exchange, we take into account a
scalar and pseudoscalar exchange in the one-loop vertex, and self-energy, 
{\em i.e.},
\begin{eqnarray}
\Lambda_{(1)}^\mu(k,p)&=&\Lambda^\mu_{(1V)}(k,p)+\Lambda^\mu_{(1S)}(k,p)
+\Lambda^\mu_{(1P)}(k,p),\label{sumvert}\\
\hat p {\cal Z}_{(1)}(p^2)&=&-\Sigma_{(1V)}(p)-\Sigma_{(1S)}(p)
-\Sigma_{(1P)}(p).\label{sumself}
\end{eqnarray}
With one-loop vertex corrections
\begin{eqnarray}
(-ie_0) \Lambda^\mu_{(1V)}(k,p)&=&
\int_\Lambda\frac{d^4 w}{(2\pi)^4}\,
(-ie_0)\gamma^\lambda iS(k-w)\nonumber\\&&\times (-ie_0)\gamma^\mu iS(p-w)
(-ie_0)\gamma^\sigma iD_{\lambda\sigma}(w),\\
(-ie_0) \Lambda^\mu_{(1S)}(k,p)&=&
\int_\Lambda\frac{d^4 w}{(2\pi)^4}\,
(-ig_Y){\bf 1} iS(k-w)\nonumber\\&&\times (-ie_0)\gamma^\mu iS(p-w)
(-ig_Y){\bf 1} i\Delta_S(w),\\
(-ie_0) \Lambda^\mu_{(1P)}(k,p)&=&
\int_\Lambda\frac{d^4 w}{(2\pi)^4}\,
(-ig_Y)i{\gamma_5} iS(k-w)\nonumber\\&&\times (-ie_0)\gamma^\mu iS(p-w)
(-ig_Y)i{\gamma_5} i\Delta_P(w),\label{pseudooneloop}
\end{eqnarray}
and the self-energies
\begin{eqnarray}
i\Sigma_{(1V)}(p)&=&\int_\Lambda\frac{d^4 k}{(2\pi)^4}\,
(-ie_0)\gamma^\mu iS(k) (-ie_0)\gamma^\nu iD_{\mu\nu}(k-p),\\
i\Sigma_{(1S)}(p)&=&\int_\Lambda\frac{d^4 k}{(2\pi)^4}\,
(-ig_Y){\bf 1} iS(k) (-ig_Y){\bf 1} i\Delta_S(k-p),\\
i\Sigma_{(1P)}(p)&=&\int_\Lambda\frac{d^4 k}{(2\pi)^4}\,
(-ig_Y)i{\gamma_5} iS(k) (-ig_Y) i{\gamma_5} i\Delta_P(k-p).
\end{eqnarray}
Taking free massless fermion, scalar, and pseudoscalar propagators,
and  the photon propagator in the Feynman gauge ($a=1$),
\begin{eqnarray}
S(p)=\frac{\hat p}{p^2},\qquad
D_{\mu\nu}(q)=-\frac{g_{\mu\nu}}{q^2},\qquad 
\Delta_S(q)=\Delta_P(q)=\frac{1}{q^2},
\end{eqnarray}
the one-loop vertices can be expressed as
\begin{eqnarray}
\Lambda^\mu_{(1V)}(k,p)&=&2e_0^2R^\mu(k,p)+2e_0^2 S^\mu(k,p),\\
\Lambda^\mu_{(1S)}(k,p)&=&-g_Y^2 R^\mu(k,p)+g_Y^2 S^\mu(k,p),\\
\Lambda^\mu_{(1P)}(k,p)&=&\Lambda^\mu_{(1S)}(k,p),
\end{eqnarray}
where the last identity is obtained form Eq.~(\ref{pseudooneloop}) 
by using ${\gamma_5} \gamma^\mu=-\gamma^\mu {\gamma_5}$, and where
\begin{eqnarray}
R^\mu(k,p)&\equiv&-i\int_\Lambda\frac{d^4w}{(2\pi)^4}\,
\frac{\gamma^\mu(\hat p-\hat w)(\hat k-\hat w)/2
-(\hat k-\hat w)(\hat p-\hat w)\gamma^\mu/2 }{
(k-w)^2(p-w)^2 w^2},\\
S^\mu(k,p)&\equiv&-i\int_\Lambda\frac{d^4w}{(2\pi)^4}\,
\biggr[
\frac{(k-w)\cdot(p-w)\gamma^\mu}{(k-w)^2(p-w)^2 w^2}
-\frac{(k-w)^\mu(\hat p-\hat w)}{(k-w)^2(p-w)^2 w^2}\nonumber\\
&&\qquad
-\frac{(p-w)^\mu(\hat k-\hat w)}{(k-w)^2(p-w)^2 w^2}\biggr].
\end{eqnarray}
Thus the sum of one-loop vertex corrections, Eq.~(\ref{sumvert}), 
can be rewritten as
\begin{eqnarray}
\Lambda^\mu_{(1)}(k,p)
=2\left[e_0^2-g_Y^2\right]R^\mu(k,p)+2\left[e_0^2+g_Y^2\right] S^\mu(k,p). 
\label{vertRS}
\end{eqnarray}
The sum of self-energy contributions, Eq.~(\ref{sumself}),
can be computed straightforwardly
\begin{eqnarray}
{\cal Z}_{(1)}(p^2)=
-\frac{(e_0^2+g_Y^2)}{16\pi^2}\left[\ln\left(\frac{\Lambda^2}{-p^2}\right)
+\frac{3}{2}\right], \label{selfen1}
\end{eqnarray}
in Minkowskian formulation.

The vacuum polarization up to two-loop corrections can be expressed as 
\begin{eqnarray}
\Pi(q^2)&=&\Pi_{(1)}(q^2)+\Pi_{(2a)}(q^2)+\Pi_{(2b)}(q^2),
\end{eqnarray}
where
\begin{eqnarray}
\Pi_{(1)}(q^2)&=&
-\frac{ie_0^2}{3q^2}\int_\Lambda\frac{d^4k}{(2\pi)^4}\,
\frac{{\rm Tr}\left[ P_{\mu\nu}(q)
\gamma^\nu (\hat k+\hat q) \gamma^\mu \hat k\right]}{(k+q)^2 k^2},\\
\Pi_{(2a)}(q^2)&=&
-\frac{ie_0^2}{3q^2}\int_\Lambda\frac{d^4k}{(2\pi)^4}\,
\frac{{\rm Tr}\left[\gamma_\mu (\hat k+\hat q)\gamma^\mu \hat k 
 \right]}{(k+q)^2 k^2}
\left[{\cal Z}_{(1)}((k+q)^2)+{\cal Z}_{(1)}(k^2)\right],\label{pi2a}\\
\Pi_{(2b)}(q^2)&=&
-\frac{ie_0^2}{3q^2}\int_\Lambda\frac{d^4k}{(2\pi)^4}\,
\frac{{\rm Tr}\left[\gamma_\mu (\hat k+\hat q) 
\Lambda_{(1)}^\mu(k+q,k) \hat k\right]}{(k+q)^2 k^2}.\label{pi2b}
\end{eqnarray}

The one-loop vacuum polarization $\Pi_{(1)}$ 
can be computed straightforwardly by making use of the projector
$P_{\mu\nu}(q)=g_{\mu\nu}-4q_\mu q_\nu/q^2$, which by contraction with 
the vacuum polarization tensor eliminates 
the term in $\Pi_{\mu\nu}$ proportional to the $g_{\mu\nu}$ tensor.
With this projector the quadratically divergent contribution to 
$\Pi_{\mu\nu}$, which is an artifact of a hard-cutoff regularization,
is eliminated explicitly.\footnote{The quadratically divergent 
contribution $\Lambda^2/q^2$ is
a notorious artifact of computing vacuum polarization corrections 
in the presence of a hard cutoff
({\em i.e.}, an explicit cutoff in the momentum integrations instead
of Pauli--Villars regularization, see for a recent discussion 
Ref.~\cite{fu94}).}
The result is the well-known one-loop vacuum polarization:
\begin{eqnarray}
\Pi_{(1)}(q^2)=\frac{\alpha_0}{3\pi}\left[\ln\left(\frac{\Lambda^2}{q^2}
\right)+{\cal O}(1)\right],\label{pi1}
\end{eqnarray}
with $q^2$ the Euclidean momentum and $\alpha_0=e_0^2/4\pi$.

The sum of the one-loop vertex functions is given in terms of the functions
$R^\mu$ and $S^\mu$.
The one-loop vertex in the Feynman gauge has been computed 
in Ref.~\cite{bach80}, see also Ref.~\cite{kirepe95} for arbitrary gauge.
One can show that, with $q^2=(k-p)^2$,
\begin{eqnarray}
2 e_0^2 R^\mu(k,p)&=&\Lambda_{(1R)}^\mu(k,p),\label{ReqR}\\
2 e_0^2 S^\mu(k,p)&=&\Lambda_{(1L)}^\mu(k,p)+\Lambda_{(1I)}^\mu(k,p),
\label{SeqLI}
\end{eqnarray}
with
\begin{eqnarray}
\Lambda_{(1L)}^\mu(k,p)&\equiv&
\frac{\gamma^\mu}{2}\left[-{\cal Z}_{(1)}(k^2)-{\cal Z}_{(1)}(p^2)\right]
+\frac{(k+p)^\mu(\hat k+\hat p)}{2(k^2-p^2)}
\left[-{\cal Z}_{(1)}(k^2)+{\cal Z}_{(1)}(p^2)\right], \label{ballchiuexpr}\\
\Lambda_{(1I)}^\mu(k,p)&\equiv&
\tau_2(k^2,p^2,q^2) T^\mu_2(k,p)+\tau_3(k^2,p^2,q^2) T^\mu_3(k,p)
+\tau_6(k^2,p^2,q^2) T^\mu_6(k,p),\\
\Lambda_{(1R)}^\mu(k,p)&\equiv&\tau_8(k^2,p^2,q^2) T^\mu_8(k,p),
\end{eqnarray}
where $\Lambda_{(1L)}^\mu$ is the one-loop longitudinal
part of the vertex, and where the $\tau_i T^\mu_i$'s
are the one-loop transverse parts as defined and computed in 
Refs.~\cite{bach80,kirepe95}.
By construction, this Ball-Chiu expression for the longitudinal vertex 
$\Lambda_{(1L)}^\mu$
satisfies the WTI \cite{bach80}:
\begin{eqnarray}
q_\mu\Lambda_{(1L)}^\mu(k,p)=-\hat k {\cal Z}_{(1)}(k^2)
+\hat p {\cal Z}_{(1)}(p^2).
\end{eqnarray}
Using Eqs.~(\ref{vertRS}), (\ref{pi2b}), (\ref{ReqR}), and (\ref{SeqLI}), 
we write
\begin{eqnarray}
\Pi_{(2b)}(q^2)=\left(1+\frac{g_Y^2}{e_0^2}\right)
\left[\Pi_{(2L)}(q^2)+\Pi_{(2I)}(q^2)\right]
+\left(1-\frac{g_Y^2}{e_0^2}\right)\Pi_{(2R)}(q^2),\label{pi2bcomp}
\end{eqnarray}
where
\begin{eqnarray}
\Pi_{(2j)}(q^2)&\equiv&
-\frac{ie_0^2}{3q^2}\int_\Lambda\frac{d^4k}{(2\pi)^4}\,
\frac{{\rm Tr}\left[\gamma_\mu (\hat k+\hat q) 
\Lambda_{(1j)}^\mu(k+q,k) \hat k\right]}{(k+q)^2 k^2},\qquad 
j=L,\,I,\,R.
\label{pi2j}
\end{eqnarray}

Since the one-loop transverse vertex functions themselves are finite,
{\em i.e.}, these function are independent of the cutoff $\Lambda$, 
the leading logarithmic contributions to the vacuum polarization
result from integrations over momenta $k^2 \gg q^2$ in 
$\Pi_{(2R)}$ and $\Pi_{(2I)}$. 
These leading logarithmic contribution can be found by first
deriving the $k^2\gg q^2$ asymptotic behavior of the transverse
structure functions given in \cite{kirepe95}, 
after which the integration over angles can be performed.
In the Feynman gauge $a=1$,
the asymptotic behavior $k^2\gg q^2$ of the $\tau$'s is
\begin{eqnarray}
&&\tau_2\approx \frac{\alpha_0}{24\pi}\frac{1}{k^4},\qquad
\tau_3\approx \frac{\alpha_0}{6\pi}\frac{1}{k^2}
\ln\left(\frac{q^2}{k^2}\right)
-\frac{29}{72}\frac{\alpha_0}{\pi}\frac{1}{k^2},
\nonumber\\
&&\tau_6\approx \frac{(2k\cdot q+q^2)}{2} \frac{\alpha_0}{24\pi}\frac{1}{k^4},
\qquad \tau_8\approx -\frac{\alpha_0}{2\pi}\frac{1}{k^2},
\end{eqnarray}
where $k^2$ and $q^2$ are Minskowskian momenta.
By making use of these asymptotic expressions
the integration over angles in 
$\Pi_{(2R)}$ and $\Pi_{(2I)}$ can be performed straightforwardly,
after performing a Wick rotation.
The integrations over momenta $k^2 \geq q^2$
leads to logarithmic corrections.
The result reads (in Euclidean formulation):
\begin{eqnarray}
\Pi_{(2I)}(q^2)&=& \frac{\alpha_0^2}{\pi^2}\left[
\frac{1}{24} \ln^2\left(\frac{\Lambda^2}{q^2}\right)
+\frac{29}{144}\ln\left(\frac{\Lambda^2}{q^2}\right)
+{\cal O}(1)\right],\label{pi2i}\\
\Pi_{(2R)}(q^2)&=&
\frac{\alpha_0^2}{\pi^2}\left[ 
\frac{1}{4}\ln\left(\frac{\Lambda^2}{q^2}\right)+{\cal O}(1)\right].
\label{pi2r}
\end{eqnarray}
The logarithmic corrections of $\tau_2$ and $\tau_6$ cancel each other,
and the contributions of $\tau_3$ give rise to a $\ln^2$ term.

An analogous computation can be performed for the self-energy and 
longitudinal vertex corrections.
Due to the Ball--Chiu expression (\ref{ballchiuexpr}) for 
$\Lambda^\mu_{(1L)}$,
the contributions $\Pi_{(2a)}$ and $\Pi_{(2L)}$ depend on the 
one-loop computation of the self-energy ${\cal Z}_{(1)}$ given in
Eq.~(\ref{selfen1}).
After expanding ${\cal Z}_{(1)}((k+q)^2)$ for $k^2\gg q^2$,  
\begin{eqnarray}
{\cal Z}_{(1)}((k+q)^2)&\approx& {\cal Z}_{(1)}(k^2)
+(2k\cdot q+q^2) {\cal Z}_{(1)}^{\prime}(k^2)\nonumber\\
&+&\frac{1}{2}(2k\cdot q+q^2)^2{\cal Z}_{(1)}^{\prime\prime}(k^2)
+\frac{1}{6}(2k\cdot q+q^2)^3{\cal Z}_{(1)}^{\prime\prime\prime}(k^2),
\end{eqnarray}
and using that ${\cal Z}_{(1)}((k+q)^2) \approx{\cal Z}_{(1)}(q^2)$ 
for $q^2\gg k^2$, the angular integration can be performed, and the
logarithmic corrections can be computed.
The result is 
\begin{eqnarray}
\Pi_{(2a)}(q^2)+\left(1+\frac{\lambda_Y}{\alpha_0}\right)\Pi_{(2L)}(q^2)
&=& \frac{\alpha_0(\alpha_0+\lambda_Y)}{\pi^2}\nonumber\\
&\times&\left[-\frac{1}{24} \ln^2\left(\frac{\Lambda^2}{q^2}\right)
-\frac{29}{144}\ln\left(\frac{\Lambda^2}{q^2}\right)+{\cal O}(1)\right],
\label{pi2a2l}
\end{eqnarray}
with $\alpha_0=e_0^2/4\pi$ and $\lambda_Y=g_Y^2/4\pi$.
Thus, comparing this expression with the $\Pi_{(2I)}$ term
in Eq.~(\ref{pi2bcomp}), we see that  
the ``overlapping divergencies'' ({\em i.e.}, the $\ln^2$) 
cancel
\begin{eqnarray}
\Pi_{(2a)}(q^2)+
\left(1+\frac{\lambda_Y}{\alpha_0}\right)\left[
\Pi_{(2L)}(q^2)+\Pi_{(2I)}(q^2)\right]= 
\frac{\alpha_0\left(\alpha_0+\lambda_Y\right)}{\pi^2} {\cal O}(1).
\end{eqnarray}
Such a cancellation occurs in a similar manner in any
covariant gauge $a$.
Thus, the two-loop contribution contribution to $\Pi$ is described
solely by the part of the transverse vertex containing the 
$T^\mu_8$ tensor,\footnote{As was shown in \cite{kirepe95}, 
this particular transverse structure function $\tau_8$ does not depend 
on the gauge parameter $a$.} {\em i.e.}, $\Pi_{(2R)}$,
and, after adding all the pieces, we deduce that 
\begin{eqnarray}
\Pi(q^2)&=& \frac{\alpha_0}{2\pi}
\left(\frac{2}{3}+\frac{\alpha_0}{2\pi}-\frac{\lambda_Y}{2\pi}\right)
\ln\left(\frac{\Lambda^2}{q^2}\right)
+(\alpha_0/\pi){\cal O}(1).\label{2loopwscalar}
\end{eqnarray}

\section{The Johnson--Willey--Baker equation}\label{dervJWB}
In this appendix we derive the equation 
\begin{eqnarray}
f_{(1)}=
\frac{N\alpha_0}{\pi}
\left[\frac{2}{3}+\frac{\phi_1+\phi_2(2+\phi_2)}{1-\phi_1}+\phi_3
\right]\label{fdef1}
\end{eqnarray}
for the $f_{(1)}$ function given by Eq.~(\ref{genvacpolexpr}), 
with the functions $\phi_j$ given by Eqs.~(\ref{phi1})--(\ref{phi3}).
The derivation of Eq.~(\ref{fdef1}) was given 
(for pure  QED) by Johnson, Willey, and Baker (JWB) 
in Ref.~\cite{jowiba67}.
Since their result is formulated in terms of the BS fermion-fermion 
scattering kernel $K^{(2)}$ their method is also applicable to the GNJL model.

In order to derive the result of \cite{jowiba67}, 
the following is assumed.
\begin{itemize}
\item{The fermion wave function equals one, 
${\cal Z}=1/A=1$, in the Landau gauge.
In principle, this assumption is redundant
since the JWB result is valid in any gauge.}
\item{
Internal photon propagators are replaced by their canonical form
$\Delta(q)=1/q^2$ which is self-consistent in the neighborhood of 
an UV stable fixed point. Only a single fermion loop, 
thus a single power of $\ln \Lambda$ contributes to the 
vacuum polarization.}
\item{Translational invariance of 
naively logarithmically divergent and finite momentum space 
integrals is assumed.
Implicitly, use is made of invariance under charge-conjugation ($C$)
and parity-time ($PT$) transformations.
}
\item{Also it is assumed that we are in the scaling region of the theory, 
where the only relevant dimensionless variable is $q^2/\Lambda^2$. 
We consider short distances with respect to the IR length scale 
$\xi\sim 1/|m_\sigma|$,
thus $|m_\sigma|^2\leq q^2\ll \Lambda^2$. 
}
\end{itemize}
Hence,
\begin{eqnarray}
{\cal Z}(k^2)=1/A(k^2)=1,\qquad S(k)=\frac{\hat k}{k^2},\qquad
\Gamma^{\mu}(k,k)&=&\gamma^\mu.\label{assum1}
\end{eqnarray}
The vacuum polarization tensor is
\begin{eqnarray}
\Pi^{\mu\nu}(q)=\frac{i\alpha_0}{4\pi^3}
\int_\Lambda
d^4 k\,{\rm \tilde Tr}\left[ 
S(k+q)\Gamma^\mu(k+q,k) S(k)\gamma^\nu\right],\label{vacpoltens}
\end{eqnarray}
where ${\rm \tilde Tr}$ denotes the sum over both spinor and 
flavor indices.
Since, in the chiral symmetric phase, the $N$ fermions are degenerate,
all $N$ fermion propagators and $N$ photon-fermion vertices are degenerate.
Hence, the sum over flavor indices gives rise to a factor $N$, {\em i.e.}, 
${\rm \tilde Tr}\rightarrow N{\rm Tr}$, where ${\rm Tr}$ is the sum over 
spinor indices.

\begin{figure}[t!]
\epsfxsize=10cm
\epsffile[-40 400 330 540]{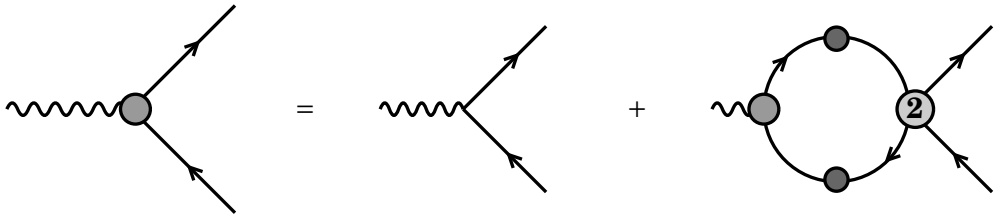}
\caption{SDE for the full photon-fermion vertex $\Gamma^\mu(k,p)$.}
\label{fig_vertex}
\end{figure}

Since vacuum polarization tensor is transverse
and the only relevant momentum variable is $q^2/\Lambda^2\ll 1$, 
the equation for the vacuum polarization can be written as
\begin{eqnarray}
\Pi(q^2)=-\frac{1}{6}\frac{q^\mu q^\nu}{q^2}
\frac{\partial^2}{\partial q_\alpha q^\alpha}\Pi_{\mu\nu}(q)
+{\cal O}(1)+{\cal O}\left( (q/\Lambda)^\sigma \right),
\end{eqnarray}
where $\sigma$ is some positive power.
After inserting Eq.~(\ref{vacpoltens}), and setting $q^2=0$ in 
the integrand, and using $q$ as the infrared cutoff in the momentum 
integral, we obtain 
\begin{eqnarray}
\Pi(q^2)&\approx&-\frac{1}{6}\frac{q_\mu q_\nu}{q^2}
\frac{iN\alpha_0}{4\pi^3}
\int_{q,\Lambda}
d^4 k\,{\rm Tr}\big[S_\alpha^\alpha(k)\Gamma^\mu(k) 
S(k)\gamma^\nu\nonumber\\
&&\qquad+2S_\alpha(k)\Gamma^{\mu,\alpha}(k) S(k)\gamma^\nu
+S(k)\Gamma^{\mu,\alpha}_\alpha(k) S(k)\gamma^\nu\big],
\label{intder1}
\end{eqnarray}
where the derivatives are defined as follows
(with $\Gamma^\mu(k)\equiv \Gamma^\mu(k,k)$):
\begin{eqnarray}
S_\alpha(k)&\equiv&\frac{\partial}{\partial k^\alpha}S(k)=
-\frac{\hat k\gamma_\alpha\hat k}{k^4},\quad
S_\alpha^\alpha(k)\equiv\frac{\partial^2}{
\partial k_\alpha\partial k^\alpha}S(k)=-\frac{4\hat k}{k^4},\\
\Gamma_{\mu,\alpha}(k)&\equiv&
\frac{\partial}{\partial q^\alpha}\Gamma_\mu(k+q,k)\biggr|_{q=0},\quad
\Gamma_{\mu,\alpha}^\alpha(k)\equiv
\frac{\partial^2}{\partial q_\alpha \partial q^\alpha}
\Gamma_\mu(k+q,k)\biggr|_{q=0}.
\end{eqnarray}
Since the integral Eq.~(\ref{intder1}) can only be proportional to 
$g^{\mu\nu}$, it reduces to
\begin{eqnarray}
\Pi(q^2)&\approx&-\frac{iN\alpha_0}{96\pi^3}
\int_{q,\Lambda}
d^4 k\,{\rm Tr}
\big[ S_\alpha^\alpha(k)\Gamma^\mu(k) S(k)\gamma_\mu\nonumber\\
&&\qquad+2S_\alpha(k)\Gamma^{\mu,\alpha}(k) S(k)\gamma_\mu
+S(k)\Gamma^{\mu,\alpha}_\alpha(k) S(k)\gamma_\mu
\big]. \label{vpbjwderv}
\end{eqnarray}
The SDE for the vertex $\Gamma^\mu$ reads, in terms of  $K^{(2)}$
(see Fig.~\ref{fig_vertex})
\begin{eqnarray}
\delta_{ij}\Gamma^{\mu}_{ab}(k+q,k)&=&
\delta_{ij}\gamma^\mu_{ab}+
ie_0^2\int_{\Lambda}\frac{d^4 p}{(2\pi)^4}\,
\left[S(p+q)\delta_{nm}\Gamma^{\mu}(p+q,p)S(p)\right]_{dc}\nonumber\\
&\times&
K^{(2)}_{^{cd,ab}_{mn,ij}}(p,p+q,k+q).\label{vdsek2}
\end{eqnarray}
From now on we omit spinor and flavor indices.
Differentiating now the SDE Eq.~(\ref{vdsek2}) 
with respect to $q$, and setting $q=0$ for the integrand and $q$ 
as IR cutoff, we obtain  
\begin{eqnarray}
\Gamma_{\mu,\alpha}(k)
&=&ie_0^2
\int_{q,\Lambda}\frac{d^4 p}{(2\pi)^4}
\biggr[S(p)\Gamma_{\mu,\alpha}(p)S(p)K^{(2)}(p,k)\nonumber\\
&+&S_\alpha(p)\Gamma_{\mu}(p)S(p)K^{(2)}(p,k)
+S(p)\Gamma_{\mu}(p)S(p)K^{(2)}_\alpha(p,k)\biggr],\label{firderv2}
\end{eqnarray}
and for the second derivative of the vertex, we find
\begin{eqnarray}
\Gamma^{\mu,\alpha}_\alpha(k)
&=&ie_0^2
\int_{q,\Lambda}\frac{d^4 p}{(2\pi)^4}
\biggr[S(p)\Gamma^{\mu,\alpha}_\alpha(p)S(p)K^{(2)}(p,k)
+2S_\alpha(p)\Gamma^{\mu,\alpha}(p)S(p)K^{(2)}(p,k)\nonumber\\
&&+S_\alpha^\alpha(p)\Gamma^{\mu}(p)S(p)K^{(2)}(p,k)
+2S(p)\Gamma^{\mu,\alpha}(p)S(p)K^{(2)}_\alpha(p,k)\nonumber\\
&&+2S^\alpha(p)\Gamma^{\mu}(p)S(p)K^{(2)}_\alpha(p,k)
+S(p)\Gamma^{\mu}(p)S(p)K^{(2)\alpha}_\alpha(p,k)\biggr],\label{secderv2}
\end{eqnarray}
where $K^{(2)}(p,k)$ and the derivatives $K^{(2)}_\alpha(p,k)$ and
$K^{(2)\alpha}_\alpha(p,k)$ are defined 
in Eqs.~(\ref{k0derdef})--(\ref{k2derdef}). 

The first derivative of the vertex is antisymmetric in $\alpha$ 
and $\mu$, because of the assumption Eq.~(\ref{assum1}).
Furthermore, $C$ and $PT$ invariance imply that the only nonzero contribution
to the first derivative of $\Gamma^\mu$ (with $q=0$ and 
$\Lambda\rightarrow\infty$) must be proportional to the tensor 
$(\gamma_\mu\hat k \gamma_\alpha-\gamma_\alpha\hat k\gamma_\mu)$.
Thus we write
\begin{eqnarray}
\Gamma_{[\mu,\alpha]}(k)&\equiv&
\Gamma_{\mu,\alpha}(k)-\Gamma_{\alpha,\mu}(k)
=\frac{(\gamma_\mu\hat k \gamma_\alpha-\gamma_\alpha\hat k\gamma_\mu)}{k^2} 
\Gamma^\prime,\\
\Gamma_{(\mu,\alpha)}(k)&\equiv&
\Gamma_{\mu,\alpha}(k)+\Gamma_{\alpha,\mu}(k)=S^{-1}_{\mu\alpha}(k),
\end{eqnarray}
where $\Gamma^\prime$ is a dimensionless scalar function.\footnote{
The function $\Gamma^\prime$ is related to the transverse structure function
$\tau_8(k^2,k^2,0)$ of Ref.~\cite{kirepe95}. At the one-loop level
$\Gamma^\prime_{(1)}=-k^2 \tau_8(k^2,k^2,0)$.}
Since $S^{-1}_{\nu\mu}(k)=0$, due to the WTI 
for the vertex and Eq.~(\ref{assum1}), we find that
\begin{eqnarray}
\Gamma_{\mu,\alpha}(k)=\frac{1}{2}\Gamma_{[\mu,\alpha]}(k)
=\frac{(\gamma_\mu\hat k \gamma_\alpha
-\gamma_\alpha\hat k\gamma_\mu)}{2k^2} \Gamma^\prime. \label{gpreq2}
\end{eqnarray}

After some algebra (taking traces over spinor and flavor indices),
we can derive from Eqs.~(\ref{firderv2}) and (\ref{gpreq2}) that
\begin{eqnarray}
\lim_{q\rightarrow 0,\Lambda\rightarrow \infty}\Gamma_{\mu,\alpha}(k)
&=&\left(\phi_1+\phi_2
+\phi_1 \lim_{q\rightarrow 0,\Lambda\rightarrow \infty}\Gamma^\prime \right)
\frac{(\gamma_\mu\hat k \gamma_\alpha-\gamma_\alpha\hat k\gamma_\mu)}{2k^2}
\quad \Longrightarrow \nonumber\\
\lim_{q\rightarrow 0,\Lambda\rightarrow \infty}\Gamma^\prime
&=&\frac{\phi_1+\phi_2}{1-\phi_1},\label{gammaprime}
\end{eqnarray}
where the $\phi_j$ functions are defined  
in Eqs.~(\ref{phi1})--(\ref{phi3}). 

Since $\Gamma^\mu(k)=\gamma^\mu$, and using Eq.~(\ref{secderv2}) 
for $\Gamma^{\mu,\alpha}_\alpha(k)$, the second derivative of 
$\Gamma^{\mu,\alpha}_\alpha$ in Eq.~(\ref{vpbjwderv}) can be eliminated.
The result is
\begin{eqnarray}
\Pi(q^2)&=&\frac{N\alpha_0}{2\pi}\big[
\sum_{n=1}^{5} I_n(q^2/\Lambda^2)+{\cal O}(q^2/\Lambda^2)\big],
\label{piifie}
\end{eqnarray}
where 
\begin{eqnarray}
I_1&\equiv& -\frac{i}{48}\int_{q,\Lambda}\frac{d^4k}{\pi^2}\,
{\rm \tilde Tr}\left[S_\alpha^\alpha(k)\gamma^\mu S(k)\gamma_\mu\right],\\
I_2&\equiv& -\frac{i}{24}\int_{q,\Lambda}\frac{d^4k}{\pi^2}\,
{\rm \tilde Tr}\left[
S_\alpha(k)\Gamma^{\mu,\alpha}(k) S(k)\gamma_\mu\right],\\
I_3&\equiv& \frac{e_0^2}{24}\int_{q,\Lambda}\frac{d^4k}{\pi^2}
\int_{q,\Lambda}\frac{d^4p}{(2\pi)^4}\,{\rm \tilde Tr}\left[
S(p)\Gamma_{\mu,\alpha}(p)S(p)
K^{(2)\alpha}(p,k) S(k)\gamma^\mu S(k) \right],\\
I_4&\equiv& \frac{e_0^2}{24}\int_{q,\Lambda}\frac{d^4k}{\pi^2}
\int_{q,\Lambda}\frac{d^4p}{(2\pi)^4}\,
{\rm \tilde Tr}\left[S_\alpha(p)\gamma_\mu S(p)
K^{(2)\alpha}(p,k)S(k)\gamma^\mu S(k) \right],\\
I_5&\equiv& \frac{e_0^2}{48}\int_{q,\Lambda}\frac{d^4k}{\pi^2}
\int_{q,\Lambda}\frac{d^4p}{(2\pi)^4}\,{\rm \tilde Tr}\left[
S(p)\gamma_\mu S(p)
K^{(2)\alpha}_\alpha(p,k)
S(k)\gamma^\mu S(k)\right].
\end{eqnarray}
Using translational invariance, $C$-$PT$ invariance, 
Eqs.~(\ref{phi1})--(\ref{phi3}) and (\ref{gammaprime}), 
we can derive that 
\begin{eqnarray}
I_1&=&\frac{2}{3}\int\limits_{q^2}^{\Lambda^2}\frac{dk^2}{k^2}
+{\cal O}(1),\quad 
I_2=\left[\frac{\phi_1+\phi_2}{1-\phi_1}\right]
\int\limits_{q^2}^{\Lambda^2}\frac{dk^2}{k^2}
+{\cal O}(1),\nonumber\\
I_3&=&\phi_2 \left[\frac{\phi_1+\phi_2}{1-\phi_1}\right]
\int\limits_{q^2}^{\Lambda^2}\frac{dk^2}{k^2}
+{\cal O}(1),
\quad 
I_4=\phi_2\int\limits_{q^2}^{\Lambda^2}\frac{dk^2}{k^2}
+{\cal O}(1),\quad I_5=\phi_3\int\limits_{q^2}^{\Lambda^2}\frac{dk^2}{k^2}
+{\cal O}(1),\label{ifies}
\end{eqnarray}
where $q^2$ is the Euclidean momentum.
Substituting Eqs.~(\ref{ifies}) in Eq.~(\ref{piifie}),
we get
\begin{eqnarray}
\Pi(q^2)&=&\frac{N\alpha_0}{2\pi}\left[\frac{2}{3}+
\frac{\phi_1+\phi_2(2+\phi_2)}{1-\phi_1}+\phi_3
\right]
\ln\frac{\Lambda^2}{q^2}+
(N\alpha_0/\pi){\cal O}(1).\label{jwbres}
\end{eqnarray}
With Eq.~(\ref{genvacpolexpr}), we obtain Eq.~(\ref{fdef1}).
This is the main result of Ref.~\cite{jowiba67}.
The entire derivation did not yet specify the BS kernel $K^{(2)}$. 
Therefore Eq.~(\ref{jwbres}) is applicable 
to the GNJL model as well.
\begin{figure}[t!]
\epsfxsize=10cm
\epsffile[0 400 370 540]{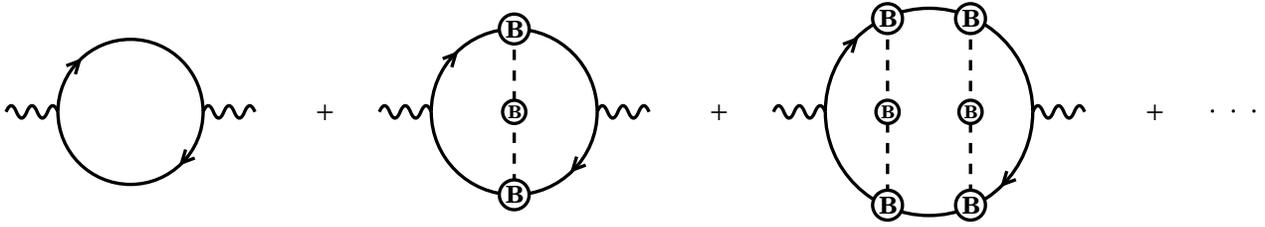}
\caption{Planar skeleton contributions to the vacuum polarization; 
the blobs represent both photon and 
$\sigma$ and $\pi$ exchanges.}
\label{fig_vacpol1overN}
\end{figure}
\end{document}